\documentstyle[12pt,fleqn,epsf,epsfig]{article}

\begin{document}

\title{On the Strength of Attractors in a High-dimensional System:
Milnor Attractor Network, Robust Global Attraction, and
Noise-induced Selection}

\author{Kunihiko Kaneko\\
{\small \sl Department of Pure and Applied Sciences, 
College of Arts and Sciences,}\\
{\small \sl University of Tokyo,}\\
{\small \sl Komaba, Meguro-ku, Tokyo 153, Japan}\\
}

\date{}
 
\maketitle

\begin{abstract}

Strength of attractor is studied by the return rate to itself
after perturbations, for a multi-attractor state of a globally coupled 
map.  It is found that fragile (Milnor) attractors have a large basin volume
at the partially ordered phase.  Such dominance of fragile attractors
is understood by robustness of global attraction in the phase space.
Change of the attractor strength and
basin volume against the parameter and size are studied.  
In the partially ordered phase,
the dynamics is often described as Milnor attractor network,
which leads to a new interpretation of chaotic itinerancy.
Noise-induced selection of fragile attractors is found that has
a sharp dependence on the noise amplitude.
Relevance of the observed results to neural dynamics and
cell differentiation is also discussed.
 
\end{abstract}
 
\pagebreak
 
 
\section{Introduction}

Study of a multi-attractor system is important in
a variety of physical, chemical, biological, and engineering problems.
In a system with many degrees of freedom, coexistence of
many attractors is rather common.  On the other hand,
memory storage to each attractor is often discussed
in application of dynamical systems to information processing,
where the abundance in attractors is required.
In neural dynamics, attribution of attractors to memory is
often adopted.  In a dynamical system model for cell differentiation,
different cell types are often regarded as different attractors in
a genetic network.

There are several approaches to a multi-attractor system.
Even for a system with low degrees of freedom,
dynamical systems studies
have revealed (fractal) basin structures and their 
metamorphose \cite{GOY}.
If the dynamics is of an overdamped type with a function to be minimized, 
the basin volume is
understood from the valley structure of such (energy) function,
as shown schematically in Fig.1a).
In this case, the dynamics can be understood through the landscape
structure, and {\sl static} representation is possible.
Indeed, for a system with many degrees of freedoms, 
rugged landscape structure has been studied in spin glass, Boolean net, and
neural networks, where {\sl static} aspects 
of a multi-attractor system are studied \cite{SG}.

When the system does not have such damping term, further
information on the phase space structure is required than the landscape.
In a high-dimensional Hamiltonian dynamical system, there are several 
attempts to make direct ``anatomy" 
\cite{Konishi,Shinjo} of the phase space.  Connection path among several
ordered states has been studied therein.

If the dynamics is high-dimensional and dissipative, but not of the 
over-damped relaxation-type,   
the structure of phase space remains totally unclear.
Even for dynamical systems with few degrees,  
the basin structure is often 
riddled \cite{riddle,Lai}, where 
the selection of attractors can be regarded almost probabilistic
for an initial point distant from the attractors.  
As the degrees get larger, the study is more difficult,
although there are some attempts in cellular automata
\cite{KK-CA} and in globally coupled map \cite{KK-GCM,Lai}.

It is necessary to distinguish the basin volume and the
stability of an attractor.  The former characterizes how large the
area for the attraction is, while the latter gives how strong
the attraction is.
There can be several possibilities on the definition of 
stability of an attractor.  In \S 2 we will introduce a few of them,
and discuss one of them in detail.  

As a specific example of a multi-attractor system, we choose
a globally coupled map (GCM), where the characterization and
coding of attractors are rather straightforward.  In \S 3,
bifurcation of several attractors in GCM is surveyed, with the
emphasis on the change of basin volume.

In \S 4, the stability of attractors in GCM is characterized using the
measures introduced in \S 2.  It is found that there is a class
of attractors that globally attracts orbits, but whose orbits are
kicked away from them by any small perturbation.  Such attractor
without the stability is called Milnor attractor\cite{Milnor,AS}. 
By adopting the
quantifiers for the stability introduced in \S 2, it is shown
that global attraction is rather robust in contrast with local stability.
Appearance of Milnor attractors is due to
the discrepancy between the  global attraction and the local stability.
In \S 5, it is shown that
Milnor attractors are quite common in the partially ordered phase in GCM.
Such dominance of Milnor attractors is preserved with the increase of
system size, and is a general feature in a system with many degrees of freedom,
as is demonstrated in \S 6.  

By perturbing an attractor with a small noise, orbits can be  switched to
a different attractor.  Through this switch, connectivity matrix
among attractors is defined depending on the input noise.   
In \S 7 this connectivity is studied.  In particular, in the
partially ordered phase, the dynamics is shown to be represented by
connection network among Milnor attractors.  This dynamics over
Milnor attractors reminds us of the chaotic itinerancy, 
previously found as the itinerant dynamics over attractor ruins.  In \S 8,
the chaotic itinerancy is re-interpreted as Milnor attractor networks.

Existence of attractors with weak stability may lead us to suspect
that the orbits might not be attracted to them in the presence of noise.
In \S 9, we have studied the rate of attraction to each attractor
in the presence of noise.  In contrary to our naive expectation, 
weak, or even Milnor, attractors may attract more orbits
in the presence of noise.  This mechanism is discussed in
relation with the global attraction in the phase space.
Complicated  dependence of the attraction rate on the noise strength
is found, which reflects the complex
connection among attractors.

Relevance of our observation to biological networks is given
in \S 10, focusing on dynamic information processing in neural dynamics,
and developmental process of cell society.  Summary and discussion are
given in \S 11 (see also ref.\cite{PRL} for rapid communication).

\begin{figure}
\noindent
\hspace{-.3in}
\epsfig{file=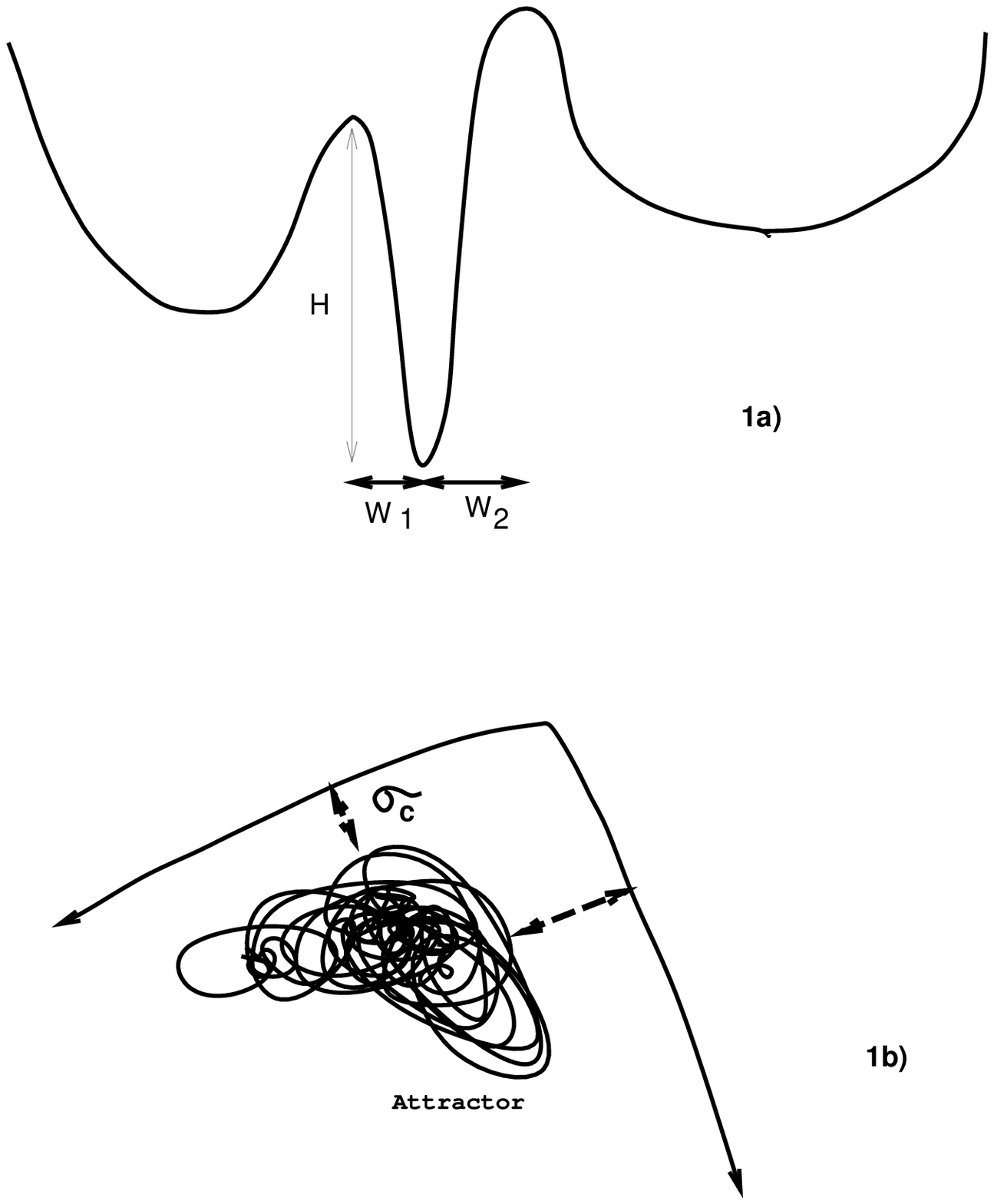,width=.75\textwidth}
\caption{Schematic representation of strength of an attractor:
a) overdamped motion of $x$ in a potential $U(x)$
b) dynamics without a potential }
\end{figure}

\section{Stability of Attractors}

It is natural to define
stability of an attractor in relation with
the degree of return of an orbit to the attractor when it is perturbed.
Now there are choices on the measure of return, and
the way how to perturb the orbit.

Let us start with a trivial example with over-damped motion
in a 'potential' (see Fig.1).  In this case, 
the stability of fixed points at the bottom of each valley can be discussed
by the depth of a hill $H$, and its basin size $W_1+W_2$.
Consider two extreme cases, one is stability against a one-shot noise
and the other for the stability under continuous presence of noise.
The stability in the former case is related with the
minimum of $(W_1,W_2)$, since
the orbit perturbed to $x_0+\delta$ remains within the same valley
if $\delta <$minimum of $(W_1,W_2)$ (see Fig.1).  On the other hand,
the dynamic stability in the presence of noise is often characterized
by the height $H$, as is typically expressed by the Kramers
formula $exp(-H/(kT))$ for the transition probability from one valley 
to another when the system is under a Gaussian white (or a similar) noise
with $kT$ corresponding to the noise strength\footnotemark.

\footnotetext{When the system is under a colored noise, it is generally
expected that not only the height but also
the shape of valley (such as its slope) is related with the
stability.}

Here we aim at generalizing the stability to a high-dimensional
dissipative case, without a potential.  To characterize an attractor,
we discuss the
basin volume, and  measures of static and dynamic stability.

The basin volume of an attractor is estimated as the ratio of initial 
points that are attracted to it, by choosing them randomly.
Even though the basin is generally not a smoothly connected object, with its
fractal or riddled basin structure, this estimate still works as 
an effective means. 

Although the basin volume ($W_1+W_2$) and the static 
stability (minimum of $(W_1,W_2)$)
are highly related in the above one-dimensional static potential case, 
they are in general independent.  
Even if the basin is smooth, the static stability is distinguishable from
a basin volume.  As a simple illustration consider the case
schematically given by Fig.1b).  As long as a one-shot perturbation is
smaller than $\sigma_c$, the orbit returns to the attractor after 
the perturbation.  For this simple case, the stability is related with the 
minimal
distance between the attractor and its basin boundary, in contrast with the
basin volume.

Although there can be several possible ways for the definition of (static)
stability, it is defined as follows here:
Perturb an orbit on an attractor at one time, and leave 
the system evolve according to its dynamics (without perturbation),
and check if the orbit comes back to it,
after transients are decayed.  The static stability is defined from
a measure of the degree of return to the original attractor.

To be specific, we define the return probability as follows.
Consider an $N$-dimensional dynamical system for $x(i)$ ($i=1,\cdots,N$).
Take an orbital point $x(i)$ on an attractor, and perturb the orbit by
$\sigma$, i.e., $x(i)+\sigma \times rnd(i)$, as $rnd(i)$ a random number
taken from $[-.5,.5]$.  
By taking this perturbed point, 
evolve the system according to the dynamics (without the noise term), and
check if the orbit reruns to the original attractor or not.
Repeat this trial a large number of times, and define the return rate
$P(\sigma)$ as the ratio of returns to the number of all trials.
$P(\sigma)$ characterizes how
paths to other attractors are opened as the orbit is perturbed from
the attractor.  The smallest $\sigma$ ($\sigma_c$) such that $P(\sigma)$ 
is less than 1
gives an estimate for the strength of an attractor.  
In the previous examples, $\sigma_c$ gives the
minimal distance between the attractor and the basin boundary.

Later we will see that there are ``attractors" with $\sigma_c =0$.
Although such ``attractors" are not asymptotically stable, we will
see that a large number of initial points is attracted to them
at some parameter regime.  For such attractors,
$P(+0) \equiv \lim_{\sigma \rightarrow 0}P(\sigma)$ gives another measure 
for the strength.

Of course,  it is often important to discuss the
form of $P(\sigma)$ itself, which is relevant to
characterize global attraction in the phase space.
As another estimate for
the strength, we define $\sigma_m$ as the half-return threshold, i.e.,
the smallest $\sigma$ such that $P(\sigma)$ is less than 0.5.

Besides the stability, it is also interesting to discuss
connection among attractors.  Using the above perturbation, one
can define the probability transition matrix $T(i,j;\sigma)$
from one attractor $i$ to another attractor $j$.  ($P_i(\sigma)=T(i,i;\sigma)$).
With this transition matrix, `connectivity' among attractors
in the presence of noise, or `distance' among attractors
in its rough sense, can be discussed. 

On the other hand, the dynamic stability can be discussed as
the residence probability to each attractor in the presence of noise. 
However, it is in general, not possible, to check precisely
if the orbit stays at an attractor in the presence of noise.  
One possible way to define the  dynamic stability is the use of the return rate
$P(\sigma)$ and the transition matrix $T(i,j;\sigma)$ 
after the noise is added over long enough time steps continuously. 
Instead of it, we check
to which attractor the orbit is settled after the noise is added for
(long enough) time steps.  The rate of attraction $V(\sigma)$ is defined
as a function of noise strength $\sigma$.  The value $V(\sigma)-V(0)$
for an attractor gives a measure of the net flow to the attractor from
others, in the presence of noise.

Some quantifiers and terms adopted in the present paper are summarized in
Table I.

\vspace{.25in}
Table I:Summary of quantifiers and terms adopted in the present paper
(see text for details)

\vspace{.1in}
\begin{tabular}{|c||c|} \hline
  Basin Volume&    the ratio of initial points attracted to the attractor,\\ 
  Ratio  $V$  &   to all the randomly chosen initial points \\ \hline
  Return Rate & Rate of points that return to the original attractor\\
   $P(\sigma)$ & after a random perturbation with the size $\sigma$ is applied 
\\ \hline
  $\sigma_c$ (strength)& the minimum strength of perturbation that leads to 
                                                    $P(\sigma)<1$ \\ \hline
   $P(+0)$        & characterizes the asymptotic stability \\ \hline
  robust attractor&   an attractor with $\sigma_c >0$ \\ \hline
  Milnor attractor& an attractor with $\sigma_c =0$, i.e., $P(+0)<1$ \\ \hline
  fragile attractor&   an attractor with $P(+0)<1$ but close to 1  \\ \hline
pseudo attractor &an attractor with $P(+0)<<1$, to be regarded as a transient\\
        & state that is trapped as an  artifact of digital computation\\ \hline
  probability transition & rate of transition form attractor $i$ to $j$ with \\
  matrix  $T(i,j,\sigma)$& the perturbation $\sigma$ \\ \hline
basin volume $V(\sigma)$ in &ratio of initial points attracted to the attractor\\
the presence of noise & with the random noise over initial time steps \\ \hline
\end{tabular}

\vspace{.2in}

\section{Revisit to partially ordered phase in GCM}

As an example of high-dimensional dynamical systems with potentiality of
many attractors, we adopt  the globally coupled map (GCM) \cite{KK-GCM}
given by
 
\begin{equation}
x_{n+1}(i)=(1-\epsilon )f(x_{n}(i))+\frac{\epsilon }{N}\sum_{j=1}^N f(x_{n}(j)),
\end{equation}

where $n$ is a discrete time step and $i$ is the index for
elements ($i= 1,2, \cdots ,N$ = system size).
Here we choose the logistic map $f(x)=1-ax^{2}$ as the local element
in eq.(1), as it has been investigated as a standard model
for a high-dimensional dynamical system.
Throughout the paper we fix the parameter $\epsilon=.1$.
In the model, attractors are known to be coded by
clustering, that is the partition of $N$ elements into mutually synchronized
clusters, i.e., a set of elements in which $x(i)=x(j)$ \cite{KK-GCM}.
Attractors in GCM are classified by the number of synchronized
clusters $k$ and the number of elements for each cluster $N_k $.
Each attractor is coded by the
clustering condition $[N_1 (\geq) ,N_2 (\geq),\cdots,(\geq )N_k )]$.
Due to the symmetry, there are at least
\begin{math} 
\frac{N!}{\prod_{i=1}^{k} N_i !} 
\prod_{over sets of N_i=N_j}\frac{1}{m_{\ell}!} 
\end{math}
attractors for each clustering condition, where $m_{\ell}$ is the
number of clusters with the same value of $N_j$.
This estimate is based on the
assumption that the cluster with the same number of elements is 
indistinguishable due to the symmetry.
For example the attractor with
$[2,2,2,2,2]$ has 945-fold degeneracy.

This number of attractors can be under-estimated, since different attractors
can exist for the
same partition.  For example, period-2 type band motion sometimes remains.
Although elements are desynchronized, they keep the relation on the
band motion ( see Fig.2 and Fig.3 for some examples of such
coexisting attractors.).
The attractors in Fig.2 can be classified by adopting band-splitting instead of
clustering.  The attractor of Fig 2a is given by (2-band;(5:5)) while that
of Fig.2b by (2-band(7:3)).  For the
attractors with the clustering of, say [3,1,1,..1], there can be a few
attractors
with a different partition into  two bands. In Fig.3a,
elements split into two bands with 4
elements clustered into [3,1] and 6 elements  mutually desynchronized,
denoted by   the band  (4[3,1],6[1,1,...,1]).
Another attractor with the same clustering [3,1,..,1] coexists 
at $a=1.65$ (and $\epsilon=0.1$), which has a band splitting to 3 and 7 elements,
where the former are synchronized and the latter desynchronized.
Since for most parameters one-to-one correspondence between clustering and
an attractor holds ( at least for that with a large enough basin volume 
to be detected numerically), and the classification both with the
band and clusterings is complicated,
we distinguish the attractors only by the clustering, unless the
use of both is necessary.

\begin{figure}
\noindent
\epsfig{file=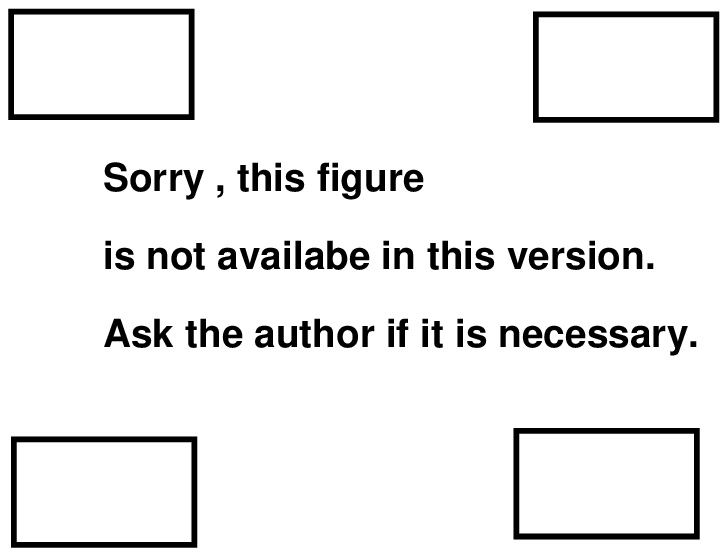,width=.8\textwidth}
\epsfig{file=nofig.ps,width=.8\textwidth}
\caption{
Overlaid time series of $x_n(i)$ of the attractor with the
clustering [1....1], accompanied by the time series of the
mean field. $a=1.66,\epsilon=.1,$ and $ N=10$.  Two examples.
(a) the band splitting with 5:5  (b) band splitting with 7:3.
}
\end{figure}


\begin{figure}
\noindent
\epsfig{file=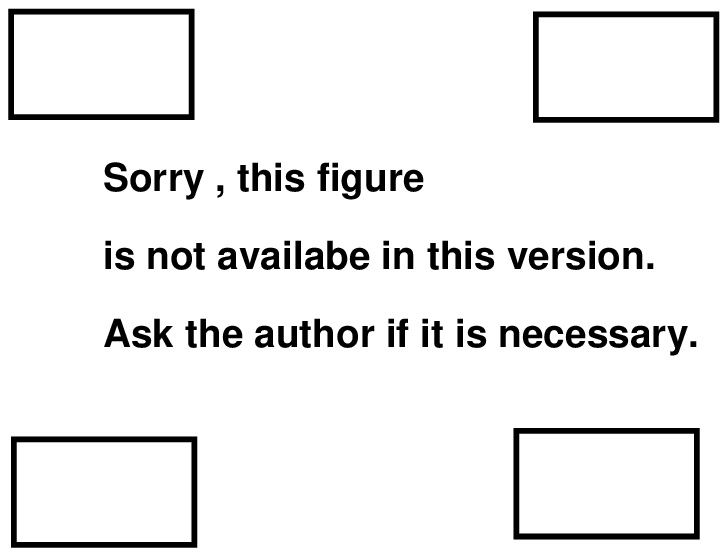,width=.8\textwidth}
\epsfig{file=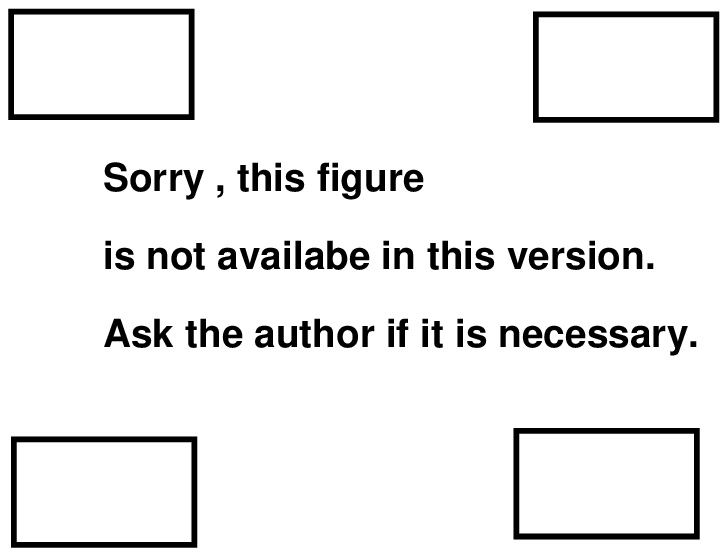,width=.8\textwidth}
\caption{
Overlaid time series of $x_n(i)$ of the attractor with the
clustering [3,1...,1], accompanied by the time series of the
mean field. $a=1.65,\epsilon=.1, N=10$.  Three examples.
(a) the band splitting with 4:6  (b) band splitting with 7:3.
}
\end{figure}


With the increase of nonlinearity $a$ or decrease of coupling $\epsilon$,
the following phases appear successively after the collapse of a completely
synchronized state;
(i) {\bf Coherent phase}: Only a coherent attractor ($k=1$) exists.
(ii) {\bf Ordered (O) phase}: All attractors consist of 
few ($k=o(N)$) clusters.
(iii) {\bf Partially ordered (PO) phase}: Attractors with a variety of 
clusterings coexist, while most of them have many clusters ($k=O(N)$).
(vi) {\bf Turbulent phase}:  Elements are completely desynchronized, and
all attractors have $N$ clusters.

In the ordered and partially ordered phases, there exists a variety of 
attractors depending on the partition. In Fig.4,
basin volumes for attractors with different clusterings are plotted,
where the decimal representation of the clusterings 
$[N_1N_2\cdots N_k]$ is adopted for each attractor.  
Successive appearance of attractors with the cluster numbers 2,3, 4 $\cdots$
proceeds at the ordered phase, and 
attractors with a large number of clusters are dominant at the PO phase.

For a system with a larger number of elements, 
enumeration of all attractors by $[N_1N_2\cdots N_k]$ is almost impossible.
Hence it is necessary to introduce some other simple measures.
The simplest measure is just the number of attractors, which 
is enhanced at the border between O and PO phases. 
A better quantifier to incorporate with the clustering
is the ratio that two elements fall into the same cluster
defined by $Y \equiv \sum_j (N_j/N)^2$ \cite{KK-part}.  
See Fig.5 for the parameter dependence of
the basin volume rate to each $Y$ value.  Note that $Y$ is close to 1/2
for a ( typical or evenly partitioned) 2-cluster attractor, 1/3 for a
typical 3-cluster attractor, and so forth.   
Successive appearance of attractors
with a larger number of  clusters is seen in Fig. 5.

The `partition complexity' is defined in \cite{KK-part} as the
fluctuation of $Y$ over initial conditions.
It is found that this fluctuation remains finite in the thermodynamic limit
at the PO phase, whose value is enhanced at the border between O and PO phases.

Another simple way is the use of entropy-like function from the
cluster probability.  The probability $p^{clust}(i)$ is defined as the basin 
ratio of
attractors with the cluster number $i$.  The function $-\sum_j p^{clust}(j) 
ln p^{clust}(j)$ has a peak again at the boundary between O and PO phases
\cite{Vul}.
Hence measures to characterize the variety of 
attractors defined from the basin volume ratio has a peak at the PO phase
(precisely speaking at the border between O and PO), while
correspondence of the
PO phase with the spin glass has been emphasized \cite{KK-part,Vul}.

\begin{figure}
\noindent
\hspace{-.1in}
\epsfig{file=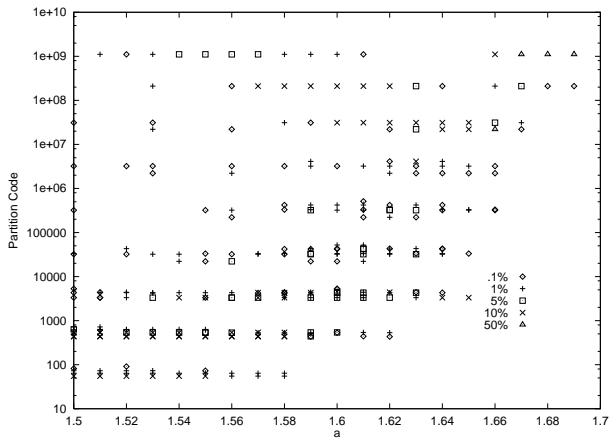,width=0.6\textwidth,angle=-90}
\caption{
Change of the partition code for existing attractors,
with the change of $a$.
The vertical axis gives the number $[N_1N_2\cdots N_k]$.  For example
the largest partition code corresponds to 1111111111, and the
smallest one is 55.  By taking
10000 initial conditions, and iterating our dynamics over 100000 steps,
we have checked on which attractor the orbit falls.  Hereafter
the basin volume rate is computed with this procedure, and
is measured as the sum of all rates over the
attractors with the same partition $[N_1,..N_k]$, unless otherwise mentioned.
The rate of initial conditions leading to such partition
is plotted as different marks.
$\triangle$( $>50$\%),$\times$ ($>10$\%),
$\Box$($> 5$\%), $+$($>1$\%), and $\Diamond$($>.1$\%).}
\end{figure}
\vspace{.2in}

\begin{figure}
\noindent
\hspace{-.3in}
\epsfig{file=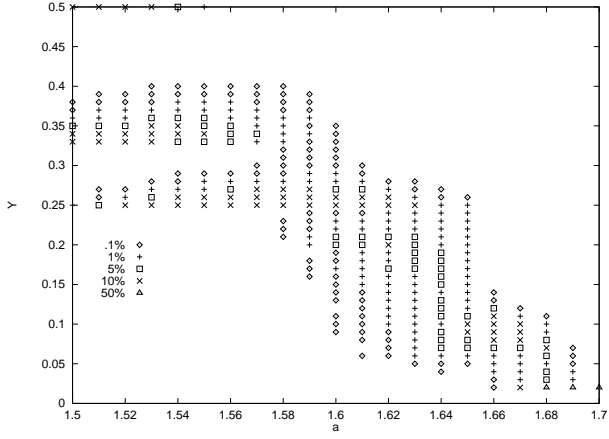,width=.6\textwidth,angle=-90}
\caption{The basin for each $Y$ value, with the change of $a$.
The rate of initial conditions leading to such value of $Y$
is plotted as different marks.
$\triangle$( $>50$\%),$\times$ ($>10$\%),
$\Box$($> 5$\%), $+$($>1$\%), and $\Diamond$($>.1$\%).
See text for the definition of $Y$.
}
\end{figure}
\vspace{.2in}




Another measure for the split of elements is given by the split exponent
$\lambda_{spl}$ defined by
\begin{equation}
\lambda_{spl}=(1/N)\sum_j \lambda_{spl}^j =(1/N)\sum_j 
log|(1-\epsilon)f'(x(j))|
\end{equation}
where the exponent gives a measure for the average split rate of two elements
taking close values\cite{KK-Inf}.

As is shown in Fig.6,
the basin volume of the attractors with $\lambda_{spl}\approx 0$
increases around the onset of partially ordered phase, while the
completely desynchronized phase appears with the further increase of $a$.
Attractors with $\lambda_{spl}> 0$ start to appear around
$a \approx 1.61$, which corresponds to the appearance of attractors with
many clusters, while the average of $\lambda_{spl}$ over 
initial conditions starts to be positive at the PO phase 
around $a\approx 1.66$.

\begin{figure}
\noindent
\hspace{-.3in}
\epsfig{file=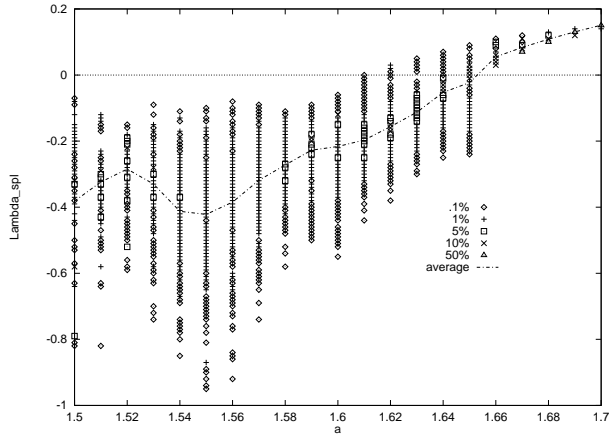,width=.6\textwidth,angle=-90}
\caption{Basin volume for attractors with $\lambda_{spl}$, as $a$ is changed.
The rate of initial conditions leading to such value of $\lambda_{spl}$
with the bin size 0.01 is plotted as different marks.
$\triangle$( $>50$\%),$\times$ ($>10$\%),
$\Box$($> 5$\%), $+$($>1$\%), and $\Diamond$($>.1$\%).
The average over 10000 initial conditions is also plotted as a line.
}
\end{figure}


Before studying the attractor strength, let us discuss the bifurcation
of the GCM in a more detail, adopting the basin volume change.
In Fig.7, the basin volume rates to attractors of given partitions
are plotted for $N=10$.  As the parameter $a$ is increased,
attractors with more clusters appear successively.
Two-cluster attractors disappear around $a \approx 1.58$,
three clusters at $a \approx 1.6$, 4, 5, and  6 clusters around 1.65.
Roughly speaking, for $a>1.65$ the attractors with the type of
$[k,1,1,1,\cdots]$ are dominant as to the basin volume.


\begin{figure}
\noindent
\hspace{-.3in}
\epsfig{file=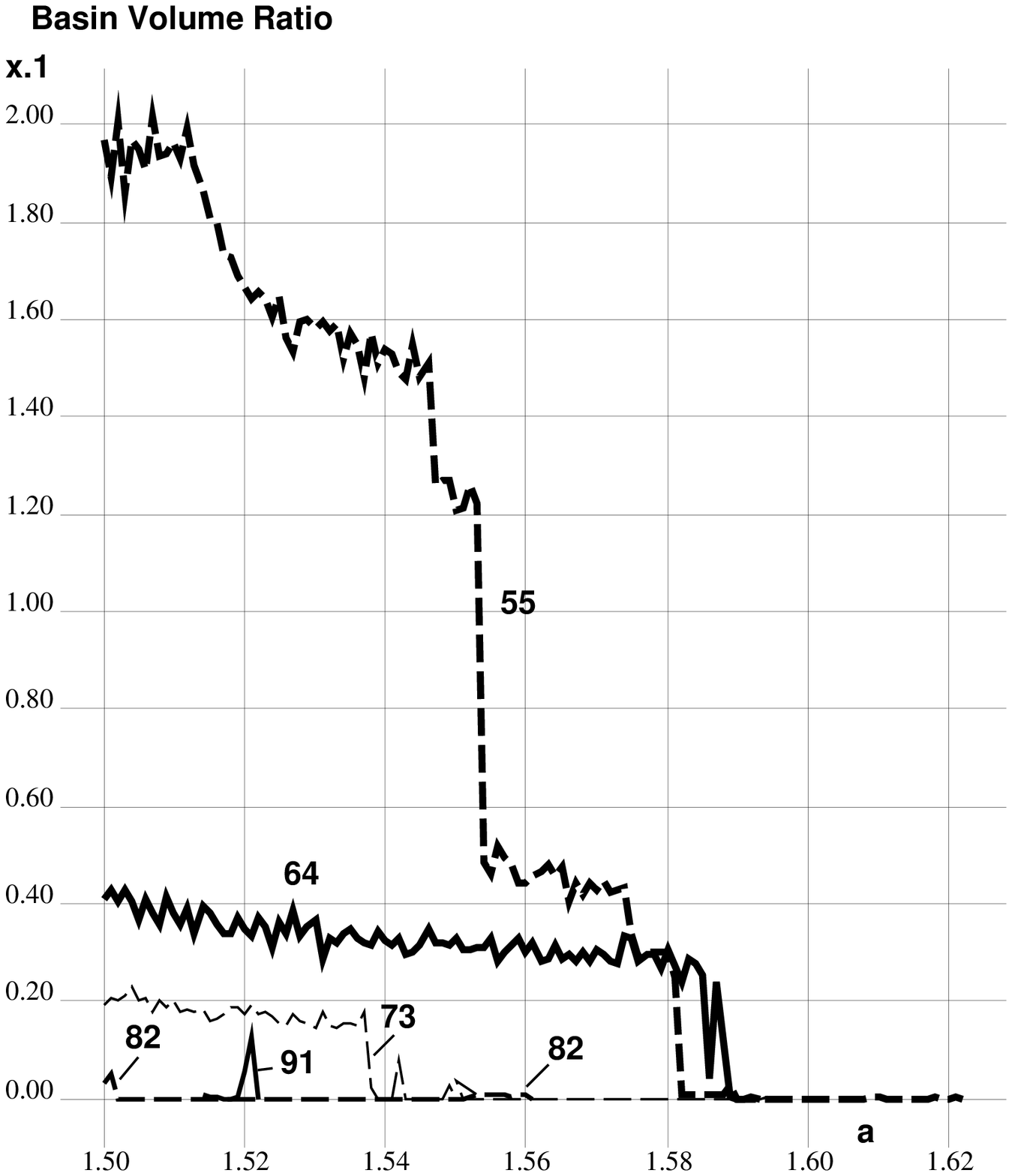,width=.4\textwidth}
\epsfig{file=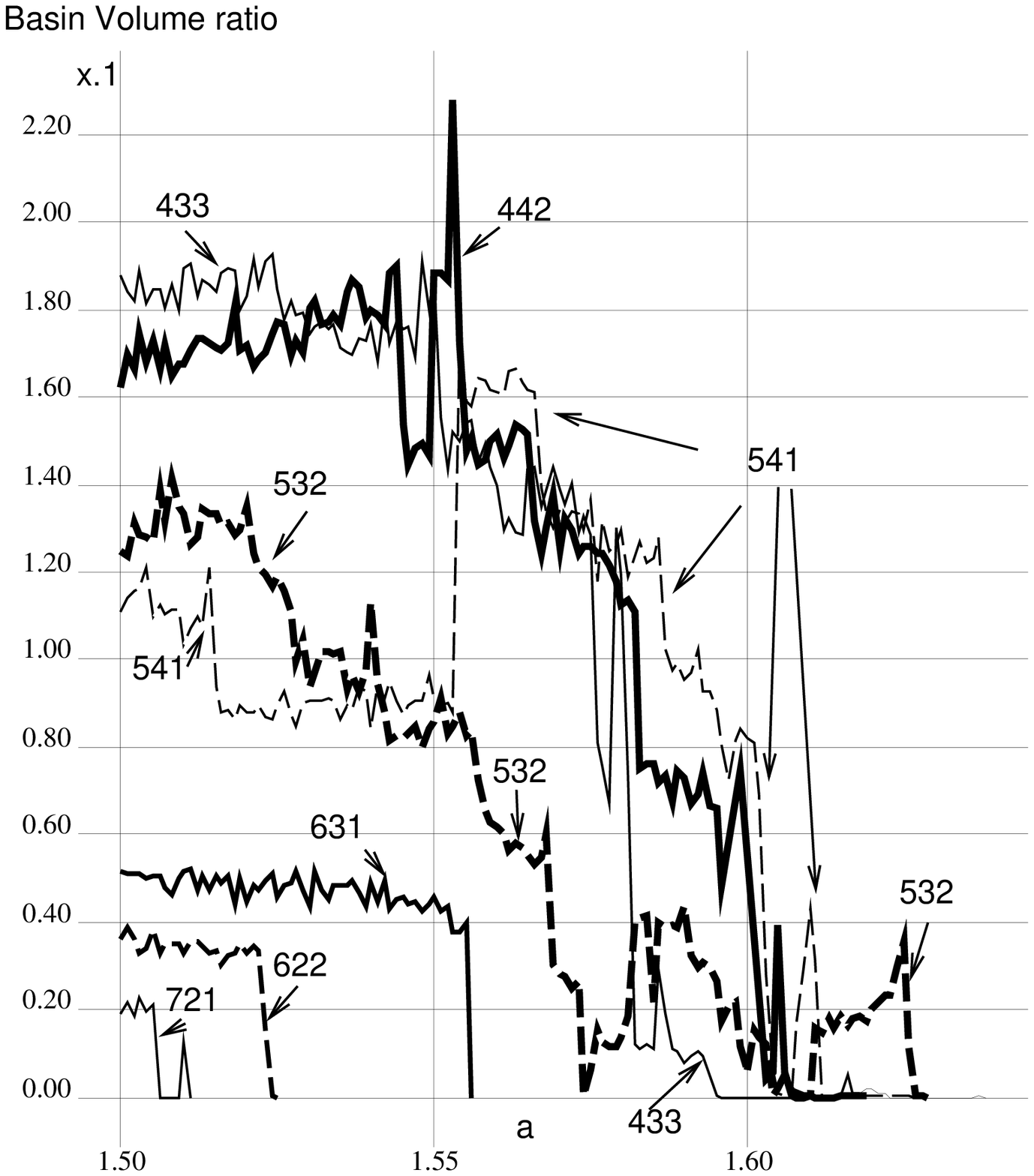,width=.4\textwidth}
\epsfig{file=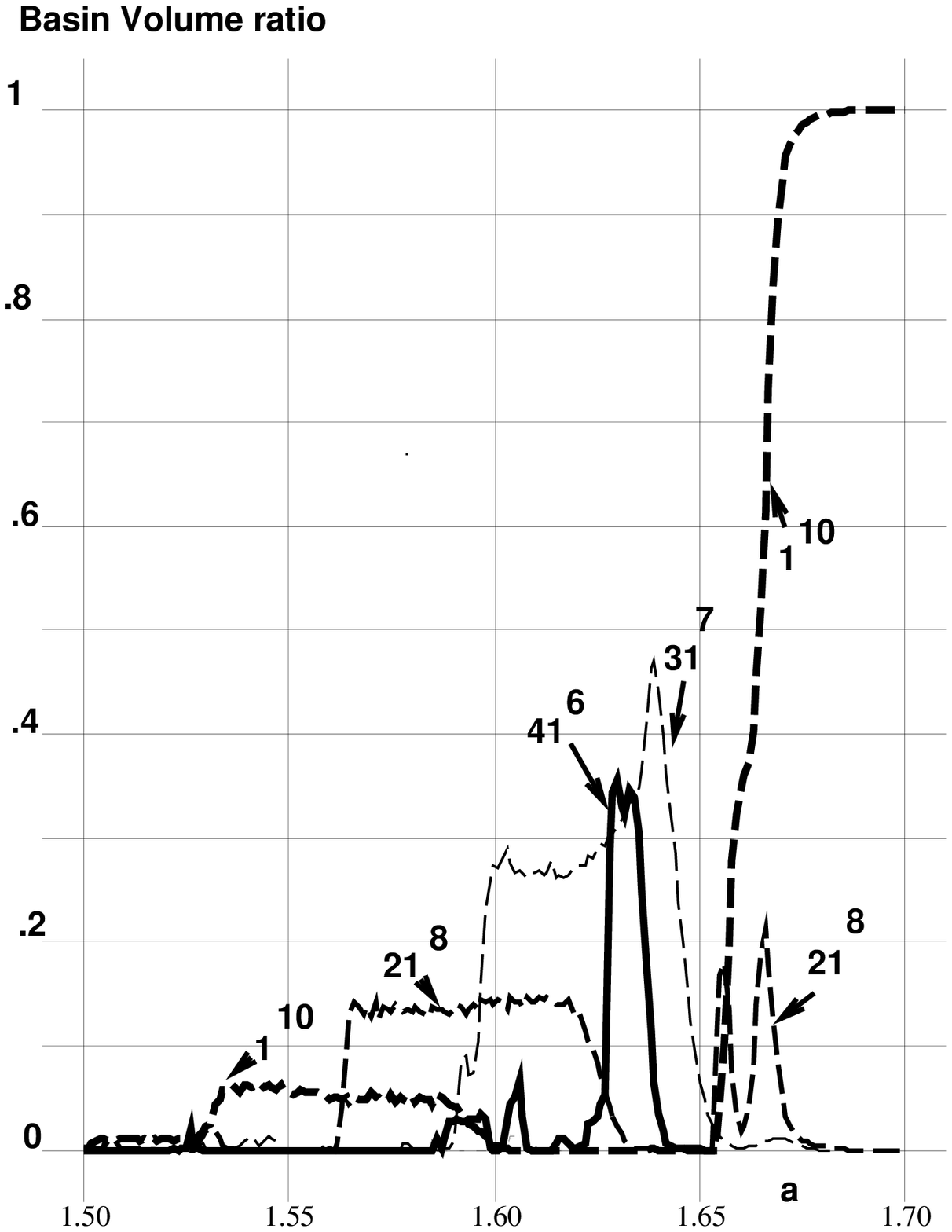,width=.4\textwidth}
\caption{Change of the basin volume rates with the parameter $a$. $N=10$.
(a) the basin volume rates for two-cluster attractors.
(b) the basin volume rates for three-cluster attractors.
(c) the basin volume rates for attractors with the partition
$[\ell,1,\cdots,1]$.}
\end{figure}
\vspace{.2in}

At the ordered phase with smaller $a$ ( e.g., $a<1.65$ for $\epsilon=.1$)
two- and three-cluster attractors are dominant.  Two-cluster attractors
lose their stability successively from biased partitions (e.g.
[9,1],[8,2],[7,3],...).   Note that there are several `cliffs' in the 
basin rate of a two-cluster attractor 
( see e.g., the basin rate for [5,5] attractors).
Indeed the cliffs are seen when an attractor with a split cluster
from it has a larger basin volume.  
For example, the cliff around $a \approx 1.55$ for a [5,5] attractor is
due to the increase of the basin volume of the [5,4,1] attractor.
By some change in the phase space structure, new paths to [5,4,1] attractors
are opened, by which some orbits previously attracted to [5,5] attractors
are attracted to [5,4,1].

As the number of clusters increases, more paths are opened or closed with
the parameter change, which leads to
complicated bifurcation structures for attractors with 4, 5, and more
clusters.  Among them, a clear structure is seen as 
successive appearance of $[\ell,1,,,1]$ attractors with decreasing $\ell$,
when $a$ is increased from 1.5, as shown in Fig. 7c).


\section{Existence of Fragile attractors and Stability of Global attraction}

Now we study how the stability of attractors change,
by using the return probability $P(\sigma)$, defined in \S 2.
See Fig.8 for examples of $P(\sigma)$ for
some attractors.  There are two types of behaviors in $P(\sigma)$.
The first one is that with $P(\sigma)=1$  up to some threshold $\sigma >0$.
Indeed, this behavior is expected for an asymptotically  stable attractor.
In this case, the `strength' of attractor is measured by
defining $\sigma_c$ as the smallest $\sigma$ such that $P(\sigma)<1$,
as mentioned.

In contrast with our naive expectation from the concept of
an attractor, there are
some `attractors' with $\sigma_c =0$, i.e.,
$P(+0) \equiv \lim_{\delta \rightarrow 0}P(\delta) <1$.
If $\sigma_c = 0$ holds for a given state, it cannot be an 
``attractor" in the sense with asymptotic stability,
since some tiny perturbations kick the orbit out of the ``attractor".
The attractors with $\sigma_c =0$
are called Milnor attractors.  In other words,
Milnor attractor is defined as an attractor that is unstable
by some perturbations of arbitrarily small size, but globally attracts
orbital points with a finite Lebesgue measure. 
Since it is not asymptotically
stable, one might, at first sight, think that it is
rather special, and appear only at a critical point like the
crisis point in the logistic map\cite{Milnor}.  Recent discovery
of riddled basin attractors, however,
leads us to expect that such attractors may not be so special\cite{GOY,Lai}.
One of the claims in the present paper
is that they are rather common at a high-dimensional dynamical system,
in particular at the partially ordered phase.

\begin{figure}
\noindent
\hspace{-.3in}
\epsfig{file=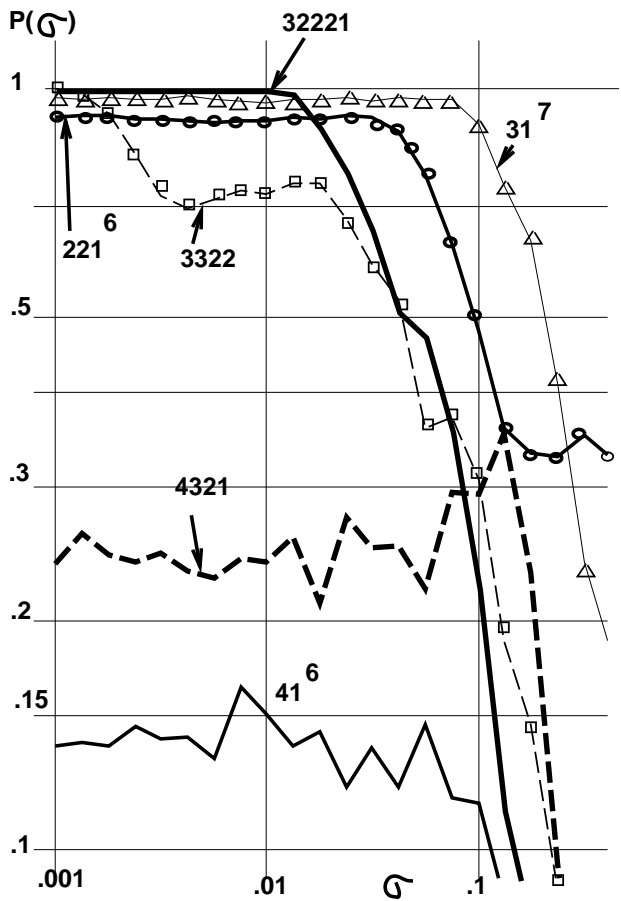,width=.6\textwidth}
\caption{
$P(\sigma)$ for several attractors
for $a=1.64$, and $N=10$.  For all the
figures,
we have 10000 initial conditions randomly chosen over $[-1,1]$
for each parameter, to make samplings.
$P(\sigma$) is estimated by sampling over 1000 possible perturbations
for each $\sigma$.
We often use the abbreviated
notation like $31^7$ for $[3,1,1,1,1,1,1,1]$.  Plotted are
robust attractors [32221] (with the basin volume rate $V \approx 6.3\%$ and
$\sigma_c \approx .01$) and
[3322] (with $V \approx 15 $\% and  $\sigma_c \approx $.0012),
fragile attractors [31111111] (with $V \approx 42 $\%), [22111111]
(with $V \approx 29$ \%), and pseudo attractors [4321] (with $V \approx.2 $\%)
and [4111111] (with $V \approx 4.8 \%$).}
\end{figure}
\vspace{.2in}


Milnor attractors of our model are rather well separated into 
two types;  one with $P(+0)$ close to 1, the other
with $P(+0)$ close to 0  (see  Fig.8).
For the former type, which we call ``fragile" attractor,
$P(\sigma)$ is close to 1 up to some perturbation strength $\sigma$ and 
decreases for larger $\sigma$.
The other type with much smaller $P(+0)$, to be called
`pseudo-attractor', shows increase in 
$P(\sigma)$ for larger $\sigma$.  (See also Table I in \S 2 for the terminology
adopted in the present paper).
One relevant index to characterize the strength of these attractors is 
given by $P(+0)$.

The above distinction between the two types of Milnor attractors
may look mathematically ill-defined.
In our simulations, however, the values of $P(+0)$ are
rather well separated, either into $>.5$ or $<.1$.
Practically speaking, we call attractors with $0.5<P(+0)<1$ as
``fragile", while those with $P(+0)<0.5$ are pseudo-attractors\footnote{
This term follows ref.\cite{KK-Inf}.}.
Indeed, the reason why we call the latter as `pseudo' is
that this attraction is thought to be due to a finite precision
in computation.  In ref.\cite{KK-Inf}, we have reported that
iterations of (1) with any finite precision
can lead to a pseudo-attractor due to artificial synchronization
(see also \cite{Okuda}).  
If the split exponent, measured over some time steps, remains negative
for long enough time, then two elements may be synchronized down to
its smallest bit in the computer.  Then, even if they are supposed to 
desynchronize later, they cannot do in a digital computer.  
Indeed, the `pseudo-attractors'
in our simulation have very small basin ratio, which can be affected by a
way of computation.  Hence we mostly focus on the
stable and fragile attractors later.

As another measure for the stability of an attractor against a
larger noise, we also use `half-decay threshold' 
$\sigma_m$ defined as the smallest $\sigma$ such that $P(\sigma)<0.5$.
For a fragile attractor $\sigma_m$ is positive, while it is zero
for a pseudo-attractor.

It is interesting to note that
$P(\sigma)$ sometimes increases with the increase of $\sigma$ (see Fig.9).
Such increase is generally seen in weak attractors, and in
Milnor attractors.  As a simple example in the ordered phase,
take a two-cluster attractor.  At $a=1.5$, we have attractors with
the clusterings
[5,5],[6,4],[7,3],and [8,2].  At this parameter regime, 
the [5,5] attractor has the largest
$\sigma_c$, which decreases as the partition is biased.
$P(\sigma)$ for the [5,5] attractor decreases monotonically, while
those for [7,3],[8,2] attractors have a double humped structure.
After $P(\sigma)$ approaches 0, it shows an increase for larger $\sigma$, 
and has an extremum around $\sigma \approx 0.5$.  Indeed $P(\sigma)$ for 
large $\sigma$
is smallest for the [5,5] attractor, and gets larger as the partition is biased.
This is an example showing that a weaker attractor can have a larger 
global attraction, which indeed is seen generally in our system.

\begin{figure}
\noindent
\epsfig{file=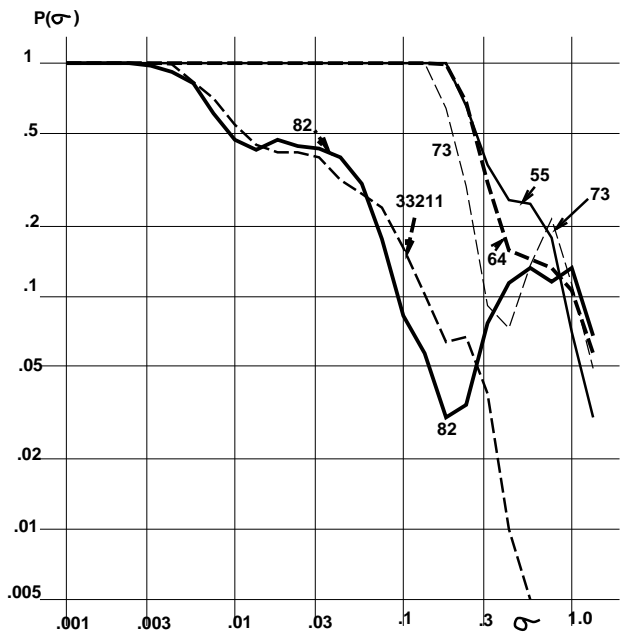,width=.6\textwidth}
\caption{
$P(\sigma)$ for two-cluster attractors, with the clustering
[5,5], [6,4] [7,3], and [8,2].}
\end{figure}
\vspace{.2in}


Extreme examples of global attraction are seen in pseudo- and fragile 
attractors.
In the pseudo-attractor, $P(\sigma)$ increases slightly as $\sigma$ is
increased from 0.
In this sense, there are some points attracted globally to the
pseudo-attractor, although there always exists a path going out of it.
In Fig.10 a)b), we have plotted such examples.

In Fig.10a), we have plotted the change of $P(\sigma)$ for the [2,2,1,1,..,1]
attractor with the increase of the parameter $a$,
as it changes from robust, to fragile, and then to pseudo attractors.
One can clearly see that the function $P(\sigma)$ 
is not much changed for large $\sigma$, although the structure
at smaller $\sigma$ is largely changed.  

Another example with a smaller number of clusters is given by
Fig.10b), where the change of $P(\sigma)$ for [5,3,2] attractor is plotted.
Although $P(0)$ changes sensitively on the parameter $a$ here, and the
attractor changes from fragile, pseudo, and robust ones.
Again, the global attraction is rather robust
in contrast with the change of stability near the attractor.

This discrepancy between  global attraction 
and local stability, as well as the stability of global attraction 
against the parameter change, will be
important later.  It should be noted that the
basin volume reflects global attraction more, which leads to
the existence of weak  or fragile attractors with a large basin volume.
Indeed, $P(0)$ that characterizes the local attraction 
often changes sensitively on the parameter $a$, but the basin volume changes
smoothly as will be shown in the next section.

\begin{figure}
\noindent
\epsfig{file=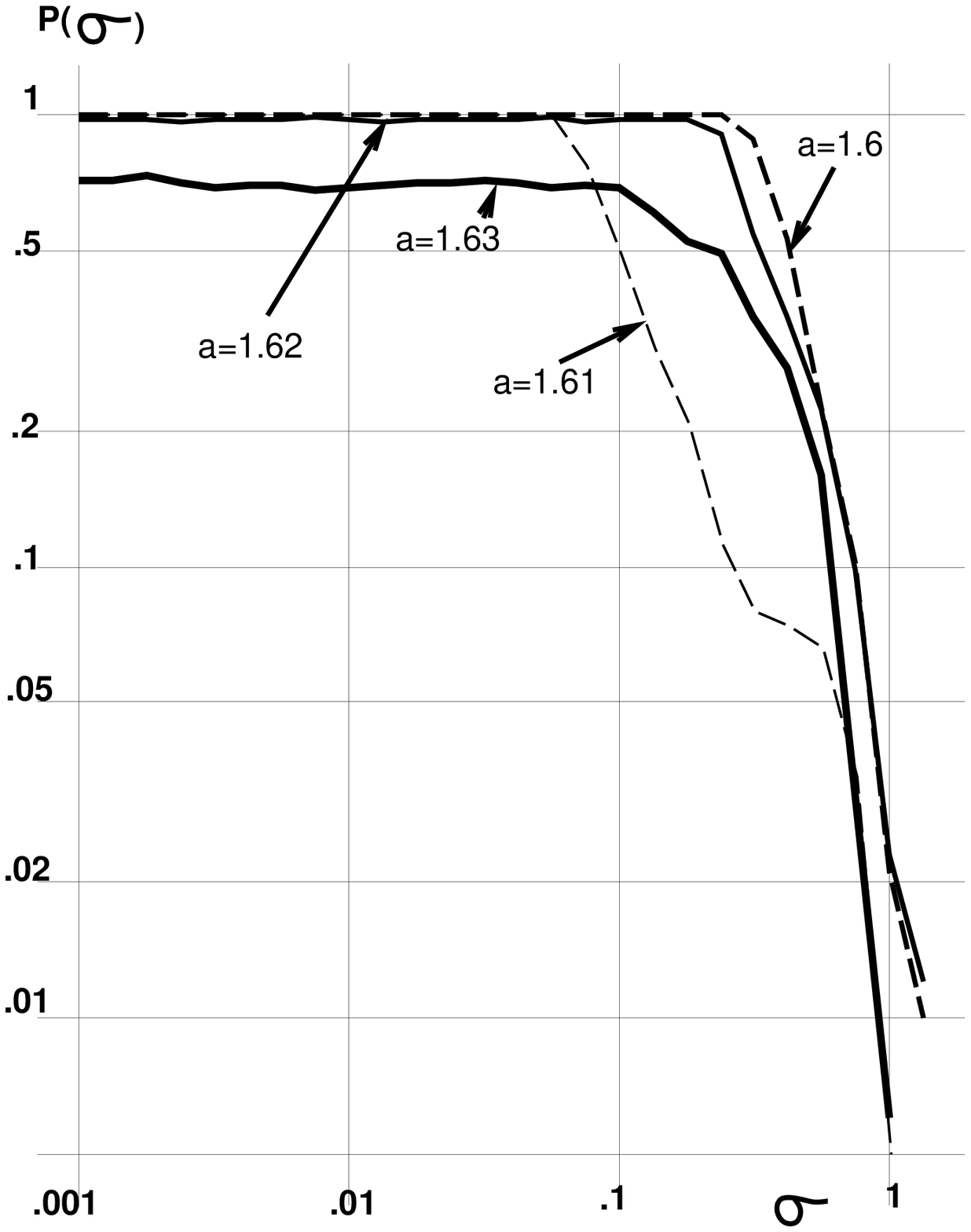,width=.6\textwidth}
\epsfig{file=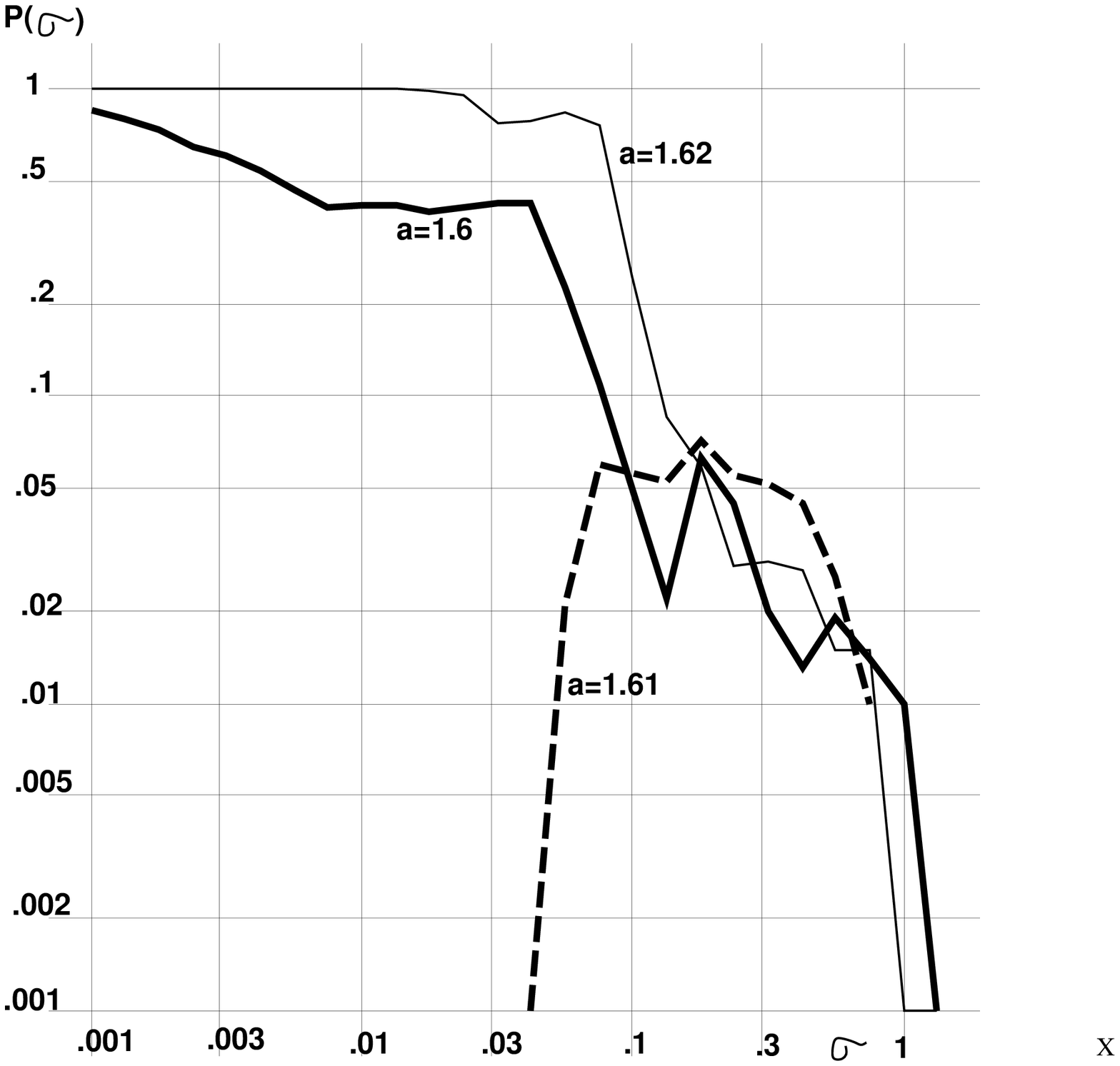,width=.6\textwidth}
\caption{
 a)   $P(\sigma)$ for [2,1,1,1,...1] with the change of $a$
for $a=1,6,1.62,\cdots 1.63$.  The attractor is robust for $a=1.6$ and 1.61,
fragile for 1.62, and pseudo for 1.63.  The inlet is the expansion
near $P(\sigma)=1$.
b) $P(\sigma)$ for [5,3,2] with the change of $a$.  The attractor is
fragile ($a=1.6$), pseudo (1.61) and robust (1.62).}
\end{figure}

%

The strength and basin volume of attractors are not necessarily correlated.
It should be noted that such fragile attractors can have a large basin volume.
Often $\sigma_c$ is small, (i.e.,the attractor is weak)
even if the basin volume is large, when
the orbit is located near the basin boundary.  In Fig.11, we have plotted
$\sigma_c$ versus basin volume rate $V$.\footnote{
As mentioned in \S 2 it is measured as the rate that 
orbits from randomly chosen initial conditions fall onto the attractor.}
Points from $a=1.57,58,\cdots 1.62$ are overlaid with different marks.
Roughly speaking, there are three  groups of attractors.
One of them keeps some relationship between the two,
($V \propto \sigma_c ^m$ with $m \approx (1 \sim 3)$), 
while two other groups are deviated from this trend.
One is a group of fragile attractors with $\sigma_c =0$ with a relatively
large basin volume and the other is
a strong attractor ($\sigma_c \stackrel{>}{\approx} 0.005$)
with relatively smaller basin volume.


\begin{figure}
\noindent
\epsfig{file=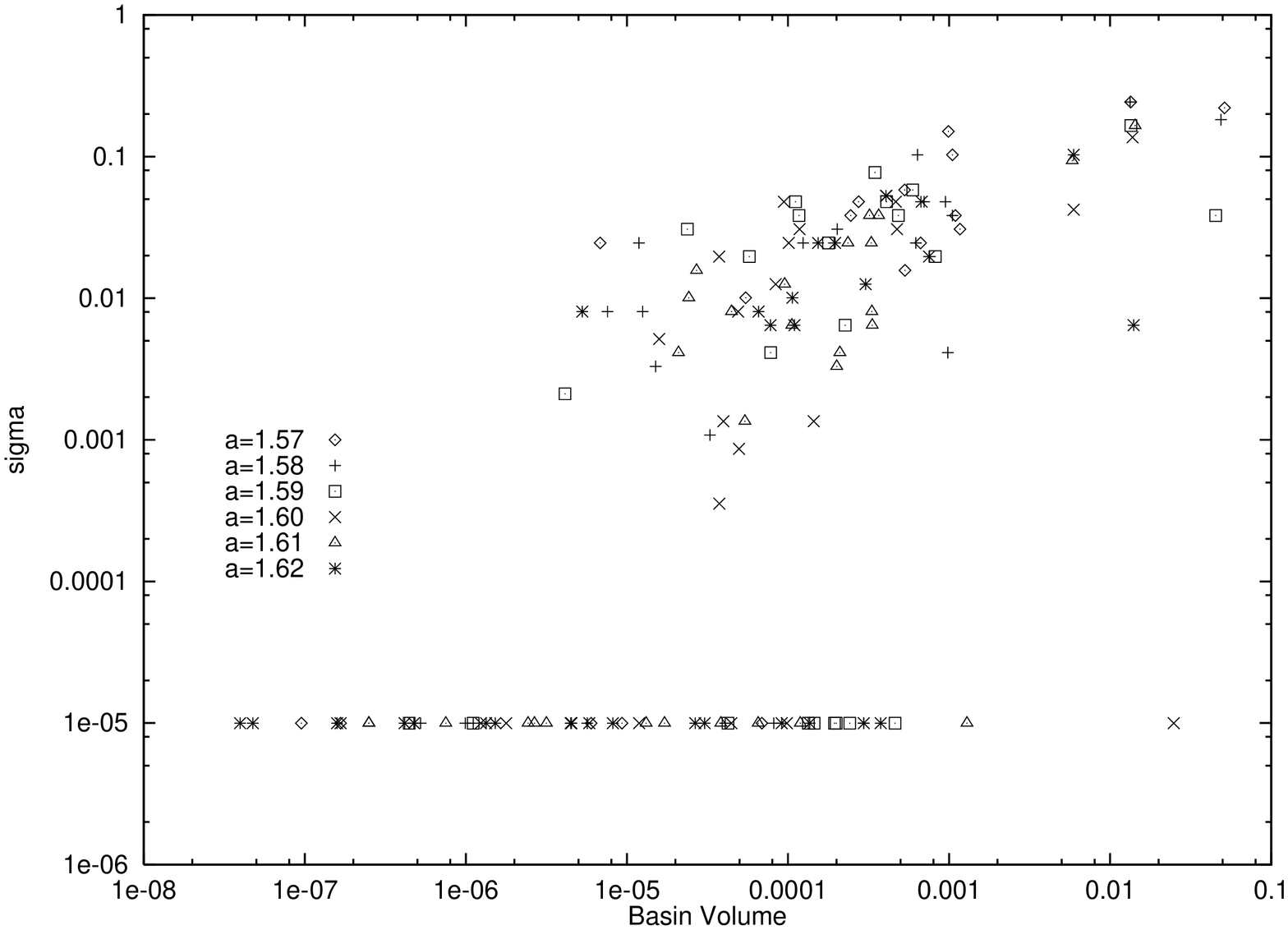,width=.6\textwidth,angle=-90}
\caption{
Strength $\sigma_c$ versus basin volume.  $\sigma_c$ is estimated
from $P(\sigma)$ measured by changing $\sigma$ 20\% successively
from $10^{-5}$.  The points at $\sigma_c =10^{-5}$ just represent that
$\sigma_c <10^{-5}$.
For Fig.11-13, we have estimated $\sigma_c$ from 100
possible perturbations: $\sigma_c$ is regarded to be less than the value
of $\sigma$ adopted in the run,
as long as all of 100 trials  result in the return to the
original attractor.}
\end{figure}
\vspace{.2in}

It may be useful to note some relationship between the basin volume and the strength
by taking a schematic example.  
If the basin is given by a hyper-ellipsoid with the radii 
$r_1(\leq)$,$r_2(\leq)$,$\cdots$,$(\leq)r_N$, and the attractor
is localized around the center of it, the strength $\sigma_c$ is given by
the minimum of $r_j$ (i.e., $r_1$).  In our clustered attractors, 
there is often some degeneracy of $r_j$ due to the symmetry. If few 
$r_j$'s ($j\leq m$) are relevant to $\sigma_c$ ($r_j \approx \sigma_ c$) 
while others remain large and insensitive to
the choice of attractors, we could roughly estimate
$V \propto \sigma_c ^m \times O(1)$.  In this context the
points on the same power-law correspond to attractors with similar 
basin shapes.

\section{Dominance of Milnor Attractors at the PO phase}




\begin{figure}
\noindent
\epsfig{file=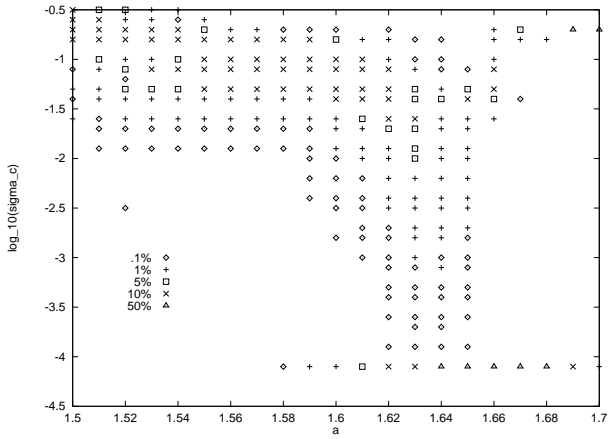,width=.6\textwidth,angle=-90}
\caption{
Dependence of $\sigma_c$ on the parameter $a$, for $N=100$.
By measuring $\sigma _c$ for attractors fallen from $10^4$ random initial
conditions, a histogram of
$log_{10}\sigma_c$ is constructed with a bin size 0.1.
The number of initial conditions leading to $log_{10}\sigma_c$ within the bin
is plotted as different marks.
$\triangle$( $>50$\%),$\times$ ( $>10$\%),
$\Box$($> 5$\%), $+$($>1$\%), and $\Diamond$($>.1$\%).
For all figures we have estimated $\sigma_c$
following the procedure given in the caption of Fig.11.
The points at $\sigma_c <10^{-4}$ just represent that
$\sigma_c <10^{-4}$.}
\end{figure}

\begin{figure}
\noindent
\epsfig{file=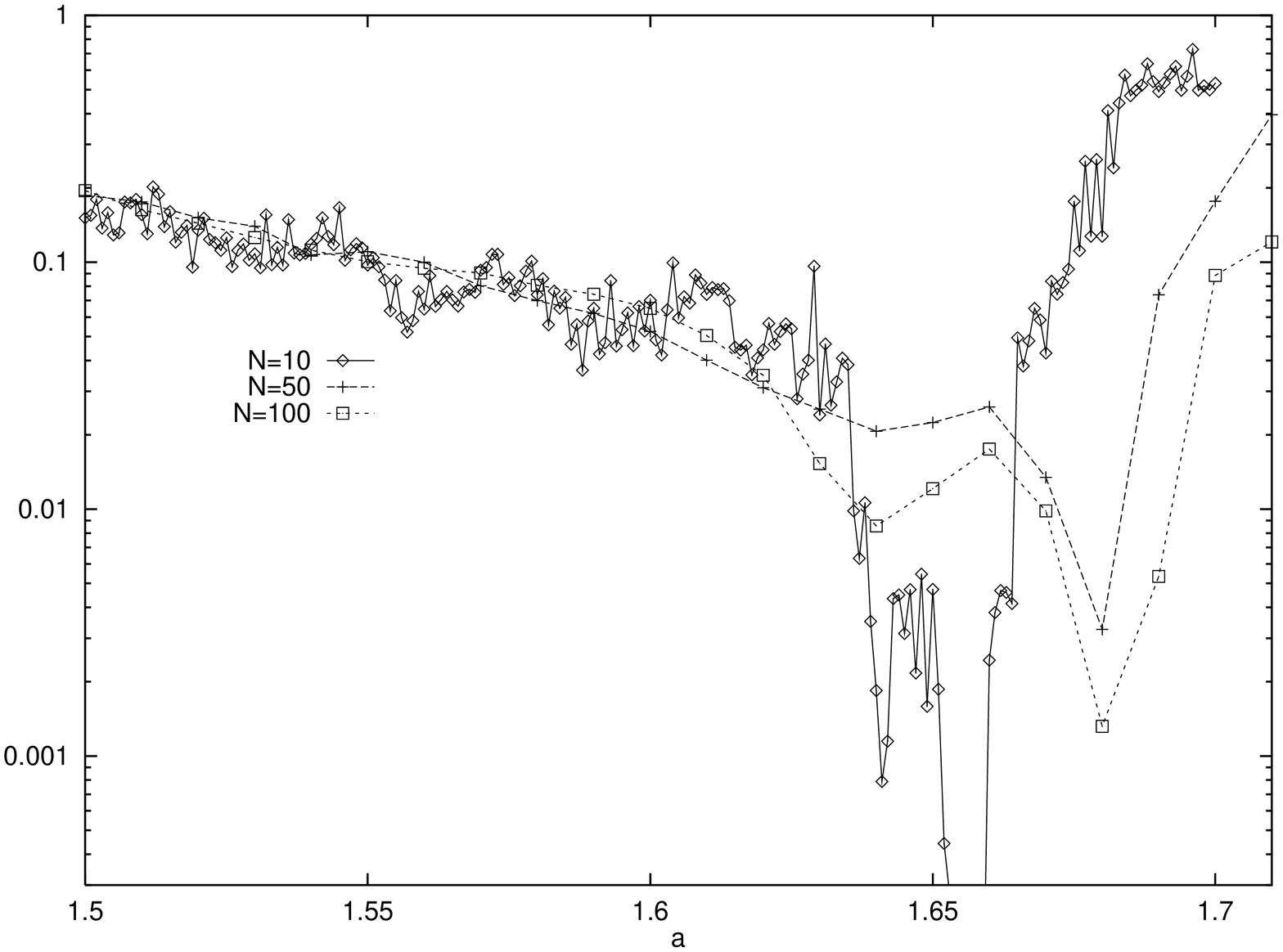,width=.6\textwidth,angle=-90}
\caption{Average strength $\sigma_c$
plotted as a function of $a$.  The basin volume of each
attractor is estimated as the rate of initial points leading to
the attractor, divided by the degeneracy.
The average is taken over $10^4$ random initial conditions.
For $N=10$, the strength is measured by increasing $a$ by 0.001,
while for $N=50$ and $N=100$ it is measured by the increment 0.01.}
\end{figure}
\vspace{.2in}

Here we study parameter dependence of the attractor strength to
see the dominance of attractors in the PO phase.
In Fig.12, we have plotted the strength $\sigma _c$ of
attractors with the change of $a$.
We note the decrease of strength at the PO phase.
The average of attractor strengths (over random initial
configurations) is plotted
in Fig.13.  The results are summarized as follows (by fixing
the parameter $\epsilon$ at 0.1).

(1) $a\stackrel{<}{\approx}1.62$\cite{boundary} ( ordered phase)
Robust attractors with 2 or 3 clusters take up a large basin volume,
although a robust attractor with $[1,1,\cdots ,1]$ with a single band (with 
a synchronized band motion) may also coexist.
No fragile attractors exist.

(2) $a\stackrel{<}{\approx}1.65$ ( complex ordered (CO) region in the ordered
phase): 
There are a variety of attractors
with different partitions, although the number of clusters is
not huge ( in other words, it is o(N) for large $N$).  The number of attractors
increases towards the border between CO and PO phases,
and the basin volume for each attractor is small\cite{PRL}.
Some attractors start to have positive $\lambda_{spl}$.
There appears fragile attractors with a large basin volume, besides
strong attractors with a small number of clusters.

(3) $a\stackrel{<}{\approx}1.68$ ( partially ordered phase):  
The split exponent 
$<\lambda_{spl}>$ averaged over initial conditions starts to be positive 
at this phase. In other words, the tendency to split elements overcomes 
the synchronization.  Thus the  number of clusters is typically large ($O(N)$),
while the basin volume for each
attractor is larger than the case (2). 
For example, for $N=10$, 
the attractors with $[2,2,1\cdots,1]$ with $P(+0)=.72$ has 60\% basin volume, 
and that with $[1,1,\cdots,1]$ with $P(+0)=.99$ has 36\%,
at $a=1.661$.  For $N=10$,
all detected attractors are Milnor attractors,
around $a=1.66$.

(4) At $a\stackrel{>}{\approx}1.69$, a single desynchronized attractor 
takes up  all basin volume.

What causes the dominance of fragile attractors at the PO phase?  
First we note that global attraction in the phase space is still kept, when
an attractor loses its stability.  This is
expected by the fact that $P(\sigma)$ at large $\sigma$ keeps a rather large
value, even when $P(+0)$  
starts to be smaller than unity.  
Recall that $P(\sigma)$ 
for large $\sigma$ is not so much changed, while  
the change from robust, fragile, to pseudo attractors proceeds. (see Fig.10).

This robustness of global attraction is a key to the understanding of
the dominance of
Milnor attractors at the PO phase.  Note there are a large number of attractors
at the border between O (CO) and PO.  Most of them lose the stability at the 
CO and PO phases successively.  When the stability of an attractor
is lost, there appears a set of points in
the vicinity of the attractor, that are kicked out of it through the temporal
evolution, while
the global attraction still remains.  This is a reason why fragile attractors 
are dominant around the PO phase.  In Fig.14, we have plotted the sum of
basin volume rates for all the Milnor attractors. Dominance of 
Milnor (fragile) attractors
is clearly seen.  It is also interesting to note that the basin rate 
for Milnor attractors sensitively depends on the parameter $a$.  

\begin{figure}
\noindent
\epsfig{file=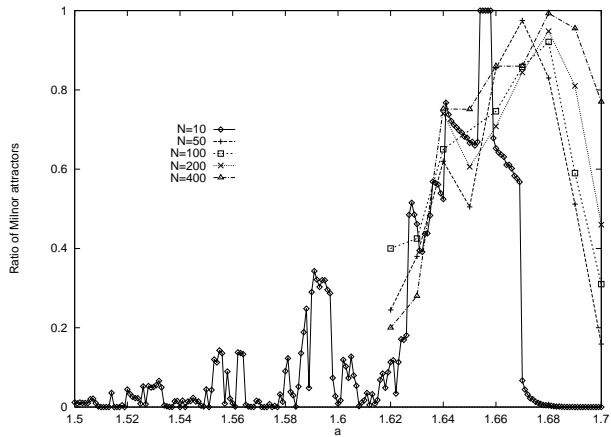,width=.6\textwidth,angle=-90}
\caption{
The basin volume ratio of Milnor attractors with the change of $a$.
For each $a$, we take 1000 initial conditions, and iterate the
dynamics over 100000 steps to get an attractor.  We check if the
orbit returns to the original attractor,
by perturbing each attractor by $\sigma=10^{-7}$ over 100 trails. If the orbit
does not return at least for one of the trails, the
attractor is counted as a Milnor one.  For $N=10$, the ratio is
measured for $1.5<a<1.7$ with the increment 0.001, while
for larger sizes it is measured only for $1.62<a<1.7$ with the
increment 0.01.}
\end{figure}
\vspace{.2in}

The results imply that attractors are often near the crisis point\cite{crisis}
and lose or gain the stability at many parameter values in the PO phase.
Furthermore, the stability of
an attractor often shows sensitive dependence on the parameter.
It is interesting to see how $P(+0)$ and basin volume change with the 
parameter $a$,
when an attractor loses asymptotic stability.  In Fig.15 we have plotted
the change of the two quantifiers for the attractors with $[3,1....,1]$,
$[2,1,...,1]$, $[2,2,1,..,1]$ and
$[1,1,...,1]$.   In Fig.15a), the basin volume has a peak when the 
attractor loses the stability and then decreases slowly as $P(0)$ gets 
smaller than unity, and the attractor becomes a Milnor one.  Although the 
local attraction
gets weaker as $P(+0)$ is smaller than 1, the global attraction remains. 
Furthermore, the basin volume often has a peak around the parameter value
of the change of stability (i.e., where $P(+0)$ starts to be less than 1),
which is rather commonly observed for several attractors.
Indeed, slight increase in the basin volume is seen for the attractors
with $[2,1,...,1]$ and $[1,1,...,1]$ when $P(0)$ gets smaller than 1.
It is also noted that
$P(0)$ often shows sensitive dependence on $a$, when it is smaller than 1.
The attractor $[2,2,1,..,1]$, with a relatively large basin volume,
is often fragile, around $1.63<a<1.67$.

\begin{figure}
\noindent
\epsfig{file=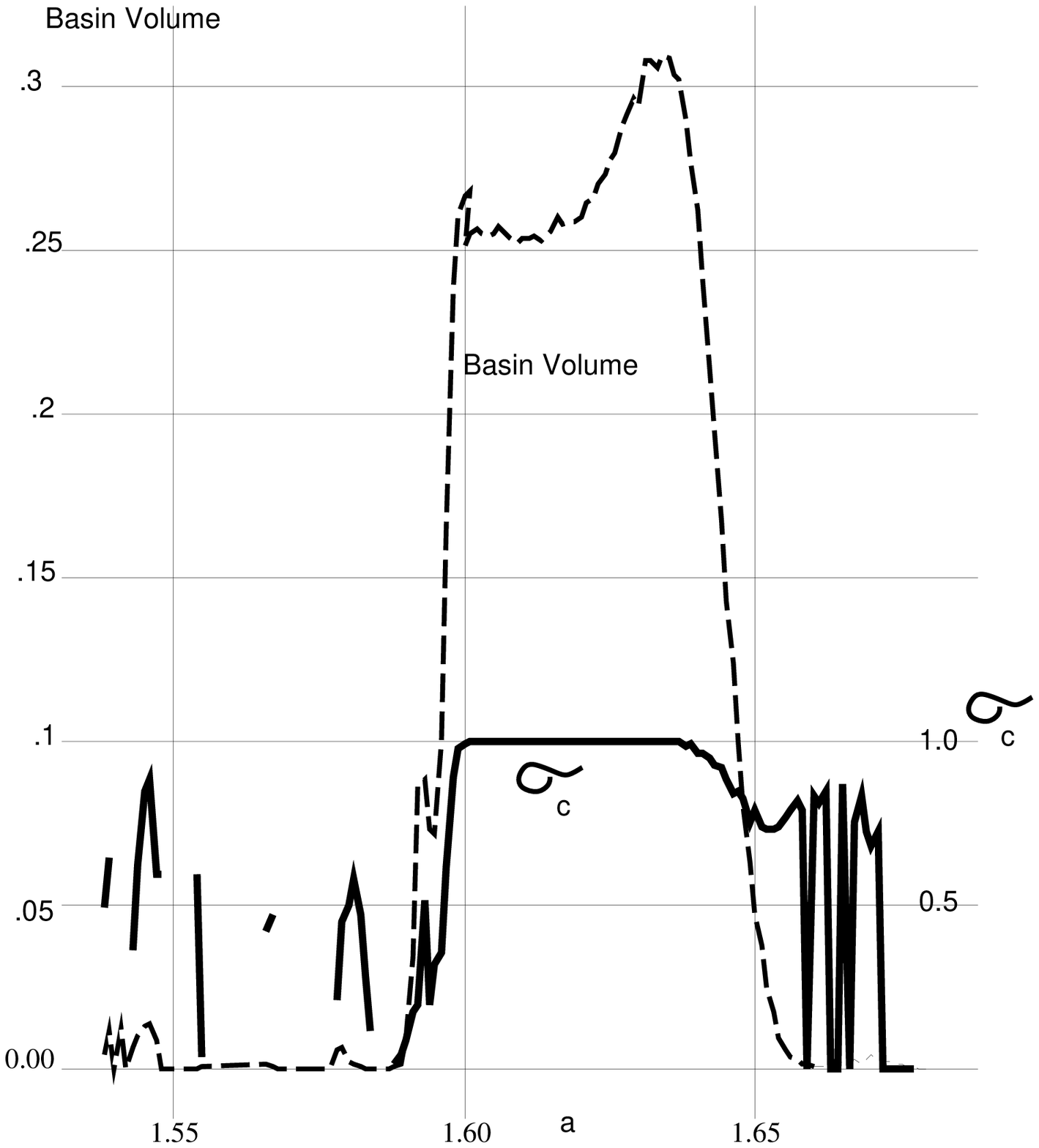,width=.4\textwidth}
\epsfig{file=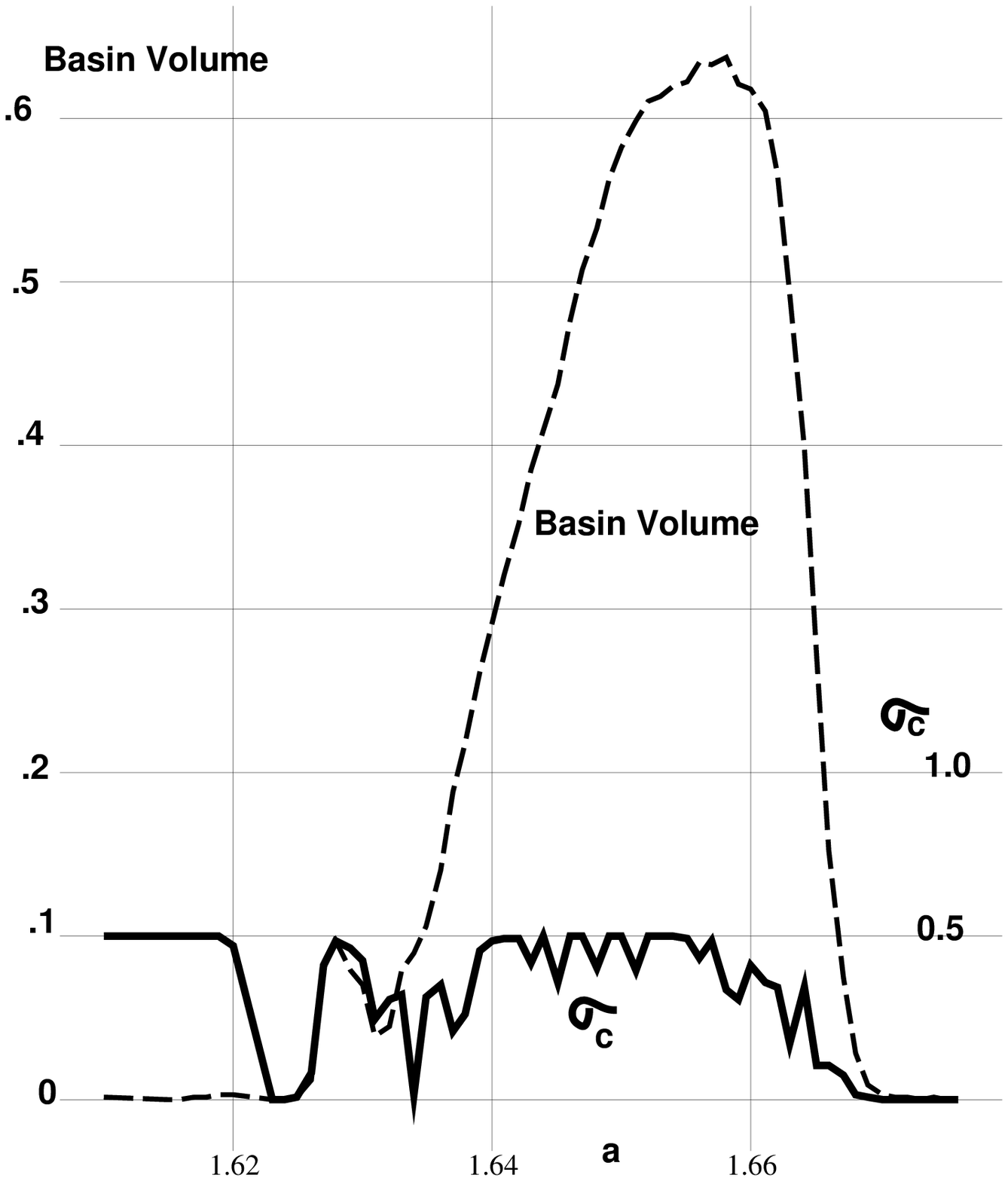,width=.4\textwidth}
\epsfig{file=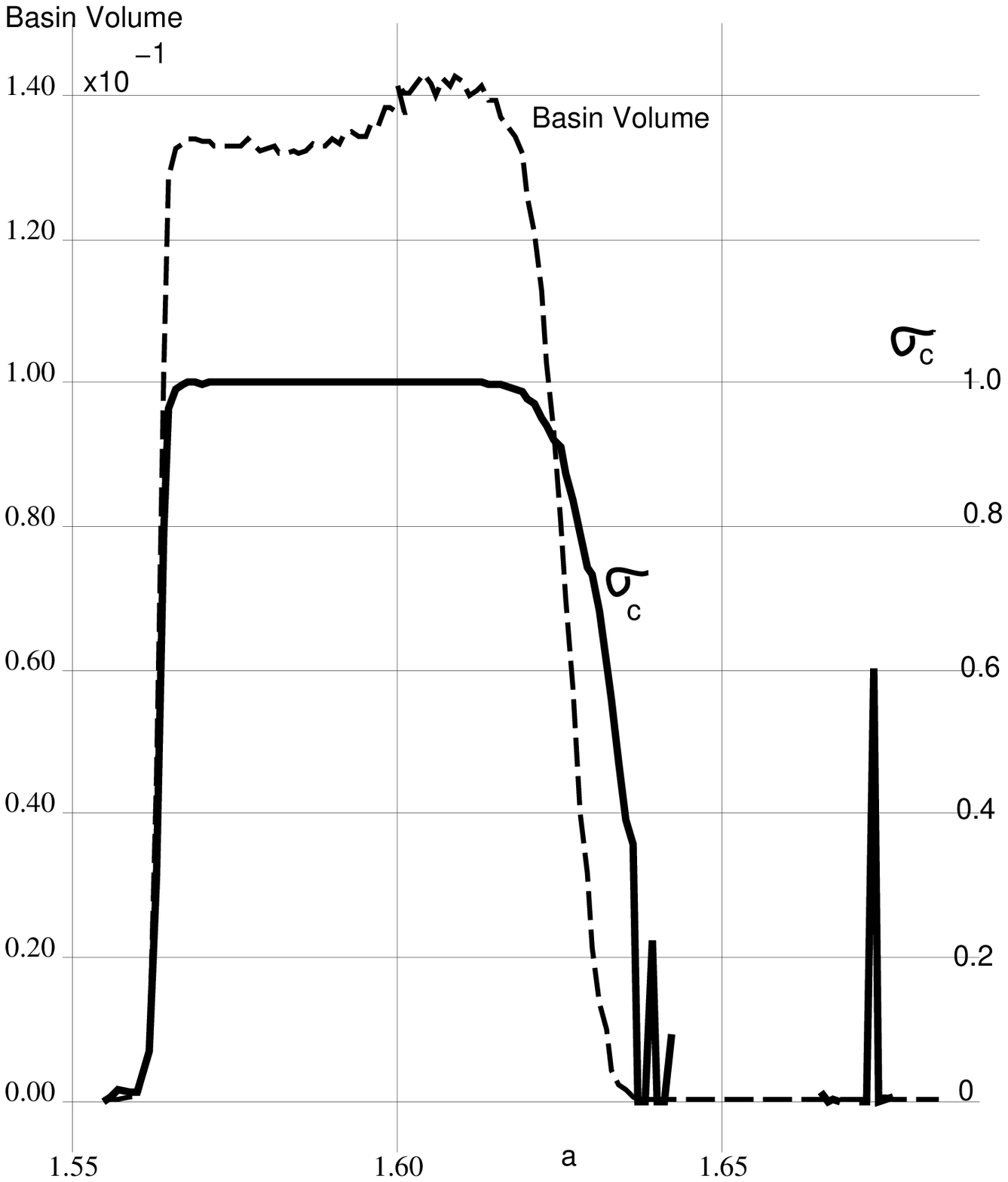,width=.4\textwidth}
\epsfig{file=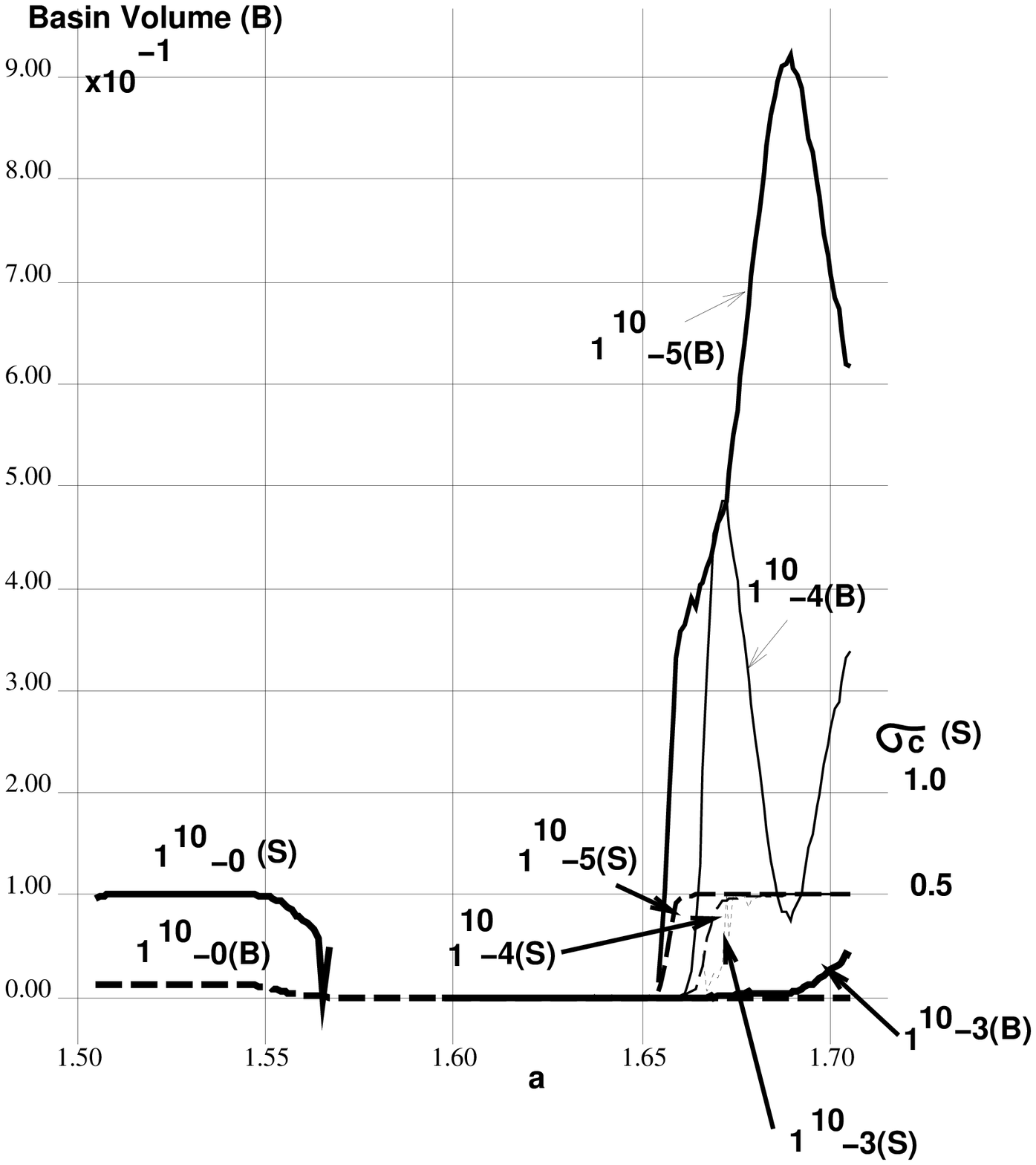,width=.4\textwidth}
\caption{
Change of the strength and basin volume rate with
the increase of $a$ by 0.001.  For each attractor,
the orbit is perturbed by $\sigma=10^{-7}$ over 1000 trails, to get
the return rate $P(+0)$, while the basin volume is measured
from $10^4$ initial conditions.  If none of the initial conditions
leads to the attractor, the strength is not plotted
(while the basin volume is plotted as 0).
(a) the attractor [3,1,..,1] with the band splitting (3:7)
(b) the attractor [2,2,1,..,1]
(c) the attractor [2,1,...,1]
(d)  the attractors [1,1,..,1] with the band splitting (10:0)
(written as $1^{10}-0$), (7:3) ( $1^{10}-3$),
(6:4) ( $1^{10}-4$), and (5:5) ( $1^{10}-5$).
"S" shows the strength $P(0)$, "B" the basin volume ratio.}
\end{figure}

\section{Size Dependence}

The dominance of fragile attractors is preserved as $N$ is increased, as
shown in Fig.14\cite{boundary}, while $<\sigma _c>$, the average of $\sigma _c$ over
initial conditions is also shown in Fig.12.
Roughly speaking, $<\sigma_c>$ seems to be smaller with 
the increase of $N$ at PO phase, although the
size dependence of $<\sigma_c>$ is irregular.  
As is discussed and seen in these figures, the PO phase shifts to a higher 
value of $a$.
Except this shift, the dominance of fragile attractors is rather common,
and  preserved 
for large $N$. 

\begin{figure}
\noindent
\epsfig{file=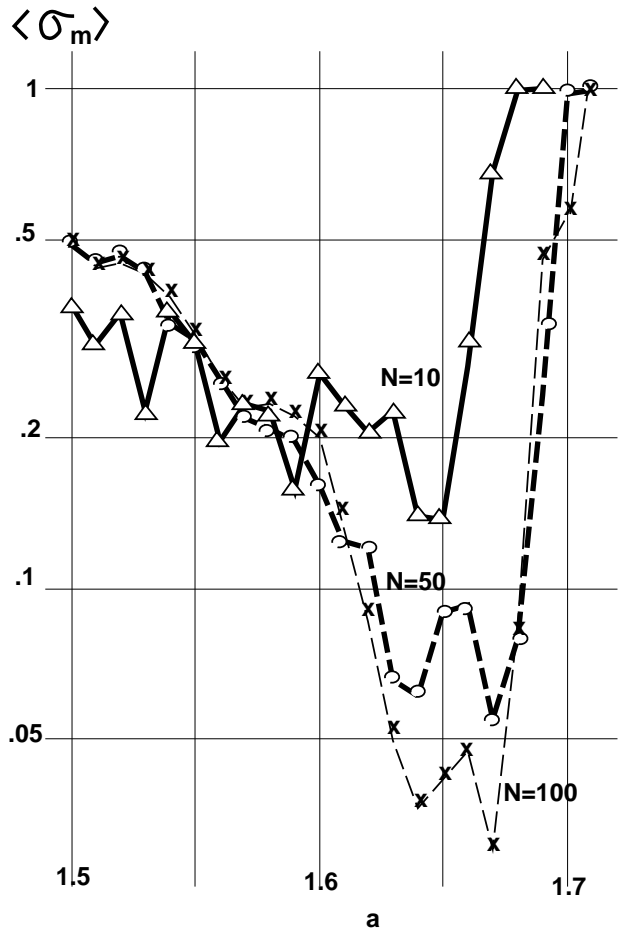,width=.6\textwidth}
\caption{
Size dependence of $<\sigma _m>$. The value
$\sigma_m$ is estimated from
$P(\sigma)$, measured by changing
$\sigma$ as $10^{-4+j/4}$ for ($j=0,1,\cdots,16$), by taking
1000 possible perturbations, while,
for $N=50$ and $N=100$, we have made only 100 possible perturbations.}
\end{figure}
\vspace{.2in}

On the other hand,
we have also plotted the average of $\sigma_m$ in Fig.16,
with the increase of the size $N$.
The average of $<\sigma _m>$ decreases with $N$, which
is related with the
escape paths from attractors.  To consider the paths,
let us discuss the size dependence of $P(\sigma)$ for each attractor.  

With the increase of $N$, the strengths $\sigma_c$ for attractors with
a proportional partition (e.g., $[6,4]$ for $N=10$ versus $[60,40]$
for $N=100$) approaches a size independent value,
while the value of $\sigma_m$ decrease monotonically.  
In Fig.17, $P(\sigma)$ for the two-cluster attractors with equal partition
(i.e., [5,5] for $N=10$ and [50,50] for $N=100$) are plotted with
the increase of the size $N$.  The strength $\sigma_c$ is invariant, while 
the slope in the decay of $P(\sigma)$ gets larger with the increase of $N$.
The latter is due to
the increase of the dimension of the path out of the attractor,
since the decrease rate in $P(\sigma)$ for
$\sigma > \sigma_c$ reflects the volume of the path.  

The dimension of the exit path is 
highly correlated with the number of elements in synchronized clusters. 
For example, take an attractor with the clustering
[3,1,1,..,1].  If the perturbation destroys the
synchronization of the first three elements, the orbit is easily
kicked out of the attractor, while stronger perturbation for the desynchronized
7 elements is required to kick the orbit.
Thus the relevant dimension of the exit path is less than
$2(=3-1)$.  Roughly speaking, the dimension of the exit path 
for small $\sigma$  is correlated with $\sum_j (N_j-1) $.  
The decrease in
$<\sigma_m>$ at the CO and PO phases (as in Fig. 16) is due to 
increase of dimensionality in
paths out of the attractors.  

In Fig.18, we have plotted $P(\sigma)$ for many-cluster attractors for
$N=50$ and $N=100$.  Here (for $a=1.65$ at the CO phase), two groups exist.
One has a larger $\sigma_c$, and positive $\lambda_{spl}$,
and a larger number of clusters with the clustering 
$[N_1,N_2,1,1,...,1]$ ( e.g., [12,10,1.,,1] for $N=50$ and $[22,21,1,...,1]$
for $N=100$).
For this group, not only the threshold $\sigma_c$ 
but also the decay slope of $P(\sigma)$ near 
$\sigma \stackrel{>}{\approx} \sigma _c$ does
not change so much  with the size $N$.
The other group has a smaller or null $\sigma_c$, negative $\lambda_{spl}$,
and a fewer number of clusters.  Examples are
[12,12,9,8,6,2,1] for $N=50$ and
[29,22,21,19,6,2,1] for $N=100$.
Although $\sigma_c$  is not changed significantly,
the decay in $P(\sigma)$ is faster with the increase of $N$,
as in the case for the two-cluster attractors.

In general, the path of exits gets larger with the size $N$
for a few number of clusters ($k = o(N)$).  The decay in $P(\sigma)$ with 
$\sigma$ is faster.  On the other hand, for an attractor with many clusters
($k = O(N)$ with $[..,1,1,..,1]$ part in the clustering),
the decay slope of $P(\sigma)$ does not change much with the size $N$.
The exit path  does not increase so much, as is expected
in the above argument for the path for $[3,1,1,1,,,,1]$.

These two distinct behaviors of $P(\sigma)$ on the size  
lead to the following implications.
First, at the PO phase, where the attractors with many clusters
(with the clustering [..,1,1,..,1] ) are dominant, the decrease of
$<\sigma_m>$ with size will stop ( see 
$<\sigma_m>$ for $N=50$ and 100  with $1.67<a<1.7$\cite{boundary}).  
On the other hand,  in the CO phase where the attractors with fewer clusters
coexist, the decrease with size continues down to the value close to
$<\sigma_c>$ ( of course $<\sigma_m>$ is bounded by $<\sigma_c>$).
As shown in  Fig.16,
the decrease is prominent near the edge of CO and PO phases, where
$<\sigma_c>$ is close to 0.

The observation that the decay slope of $P(\sigma)$ for many-cluster attractors
(with [..,1,1,..,1]) does not increase with the size
also implies that the global attraction to them 
is relatively larger than attractors with few clusters, when the
size gets larger.  In the PO phase, the attractors with  [..,1,1,..,1]
are often fragile.  The
long-term dynamics is expected to constitute in the successive 
alternations between (global) attraction to such fragile
attractors and departures from them.  This will lead to chaotic itinerancy
dynamics as will be discussed in \S 8.  The above argument implies the
importance of chaotic itinerancy for a system with a large size.

\begin{figure}
\noindent
\epsfig{file=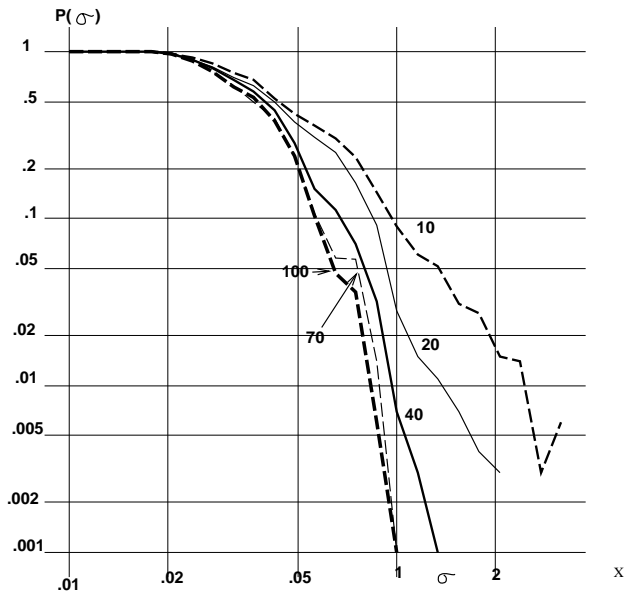,width=.6\textwidth}
\caption{
$P(\sigma)$ for two-cluster attractors with the equal
partition.  $N=10$ ([5,5]), $N=20$,[10,10], $N=40$, $N=70$, and  $N=100$
([50,50]). $a=1.5$.}
\end{figure}

\begin{figure}
\noindent
\epsfig{file=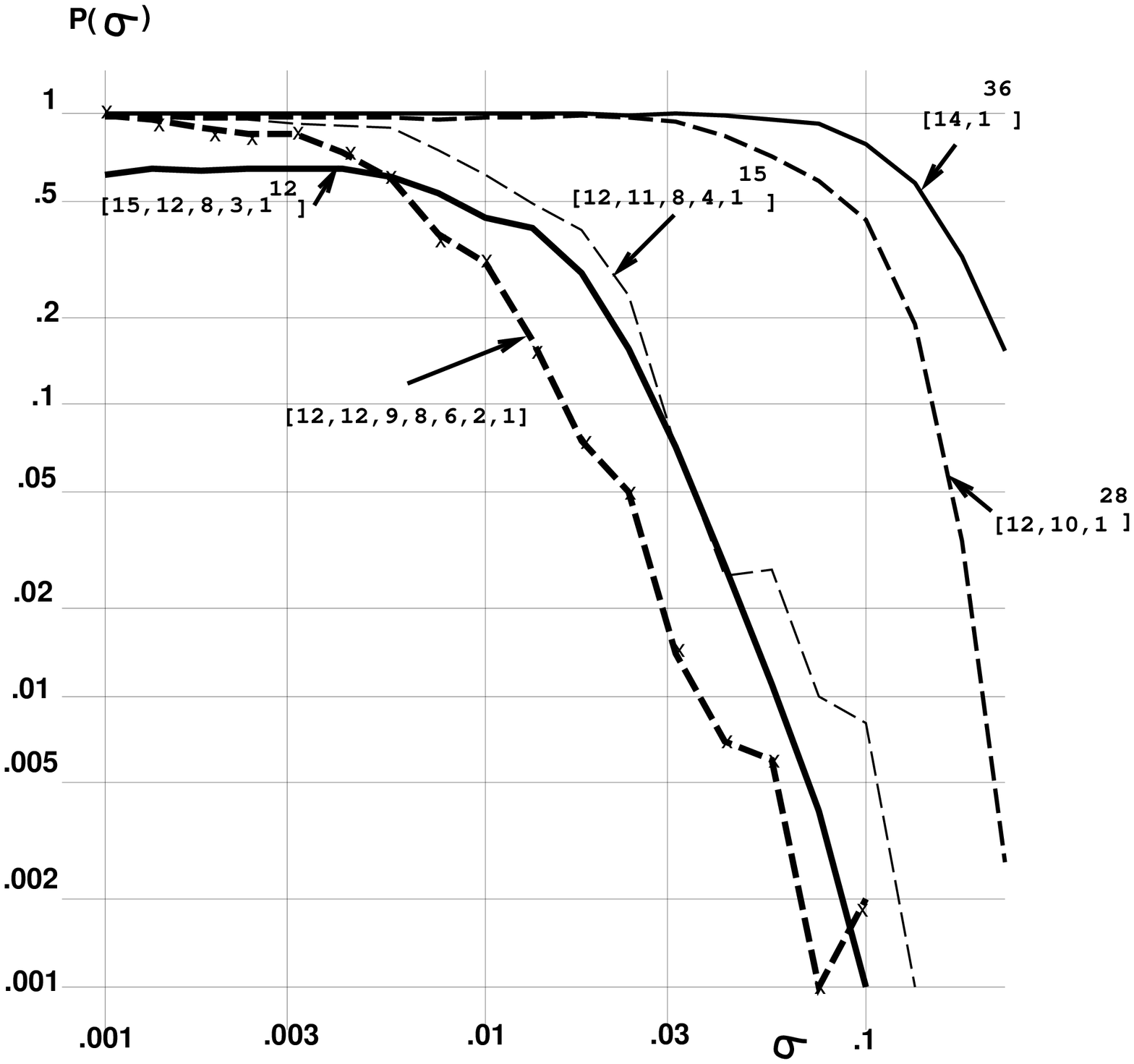,width=.4\textwidth}
\epsfig{file=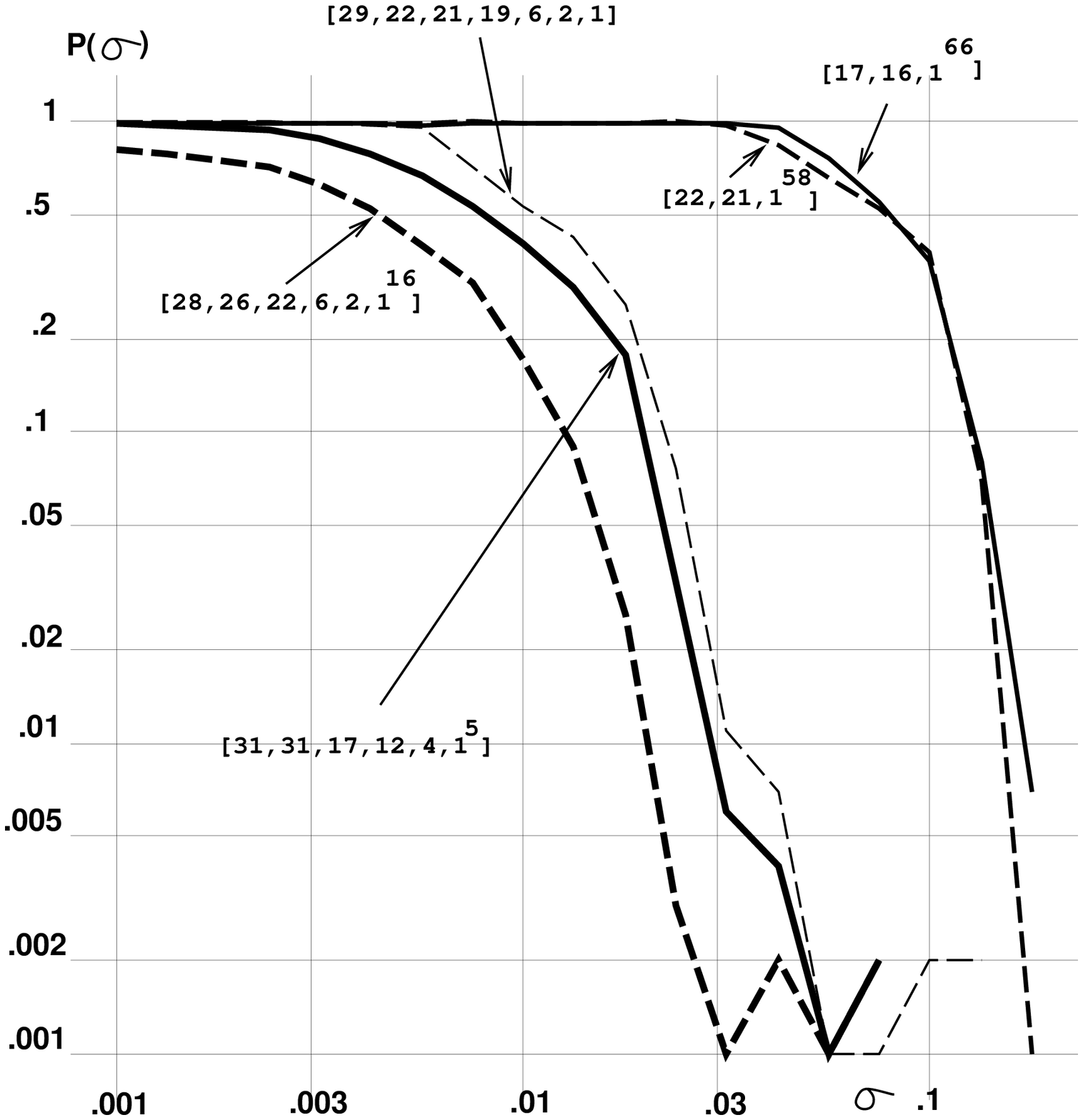,width=.4\textwidth}
\caption{
$P(\sigma)$ for several attractors with $a=1.65$.
(a) $N=50$ and  (b) $N=100$.}
\end{figure}

\section{Milnor Attractor Network}

Following the method in \S 2, 
we have also studied the transition matrix among attractors,
that give the rate of transition from one attractor to another when 
the former is perturbed by $\sigma \times rnd(i)$. In general, 
Milnor attractors are connected to a variety of attractors.  
Hence, small perturbations to such attractors make the orbit fall
into a variety of different attractors.
On the other hand, robust
attractors are mutually 
disconnected each other, and the transition between such attractors requires a
large amplitude noise.  

Typical connections that appear at a small noise are

(i)'split' ; process of $[..,n_i,...]\rightarrow [..,n_{\ell},n_m,...]$ with
$n_i=n_{\ell}+n_m$.  The simplest and most frequently observed
case is the evaporation of an element from
a cluster given by $[..,n_i,...]\rightarrow [..,n_i-1,1,...]$.

(ii) 'fusion'; process to join two clusters; the inverse process of split
$[..,n_{\ell},...,n_m,...] \rightarrow  [..,n_i,...]$with $n_i=n_{\ell}+n_m$.
The simplest and the most frequent case is the absorption of an element.

(iii) 'exchange of elements'; 
$[..,n_i,...,n_m...]\rightarrow [..,n_{i}-1,...,n_m+1,...]$.

Although these three processes are most common for small
perturbation $\sigma$, a composite process is found,
for a switch from Milnor attractors with $[...,1,1,,..,1]$.  In this case,
several elements from the $[1,1,1,..,1]$ part
join to form more than two clusters or a cluster with
more than two elements.
For example, at $a=1.63$, the transition from the
fragile attractor [2,1,...1] to [3,2,1,...,1] or [4,1,..,1] is
seen for $\sigma=+0$, although the transition matrix to
[3,1,...,1] (fusion (ii)) or to [2,1,...,1] with a
different pair of the two elements (exchange (iii)) has a larger
value.
On the other hand the switch from a robust attractor to others at 
$\sigma \approx \sigma_c$
consists of the above three fundamental processes.
For example, at $a=1.63$, the transition matrix from the
robust attractor [3,1,$\cdots$,1] to [4,1,$\cdots$,1] (fusion (ii))
starts to be positive 
first around $\sigma \stackrel{>}{\approx} \sigma_c$, and the matrix
from the  robust attractor [3,3,2,2] to [3,2,2,2,1] (split (i))
is positive around  $\sigma \stackrel{>}{\approx} \sigma_c$. 

In the limit of $\sigma \rightarrow 0$, only Milnor attractors are
connected to other attractors.  There are connections to robust attractors
if any, but mutual connections among Milnor attractors
are also observed.  Here the connection among Milnor attractors is still 
asymmetric: often, there is a
connection from fragile attractor A to fragile attractor B, but
not from B to A.

In Fig.19, we have plotted examples of the connection among
attractors in this limit.  In the complex ordered phase, 
there are  a variety of connections 
from fragile attractors to robust attractors.  Connections among
fragile attractors also exist,
which can form a network as $[2,1,...,1]\rightarrow[4,1,,...,1]\rightarrow$
$[4,2,1,.,1]\rightarrow[2,1,...,1]$ or $[2,2,...,1]\leftrightarrow[4,1,,...,1]$
as shown in Fig.17 a).   
Here it is expected that an orbit is kicked out of
Milnor attractors and is absorbed to robust attractors,
when a very small noise is continuously added to the system.

\begin{figure}
\noindent
\epsfig{file=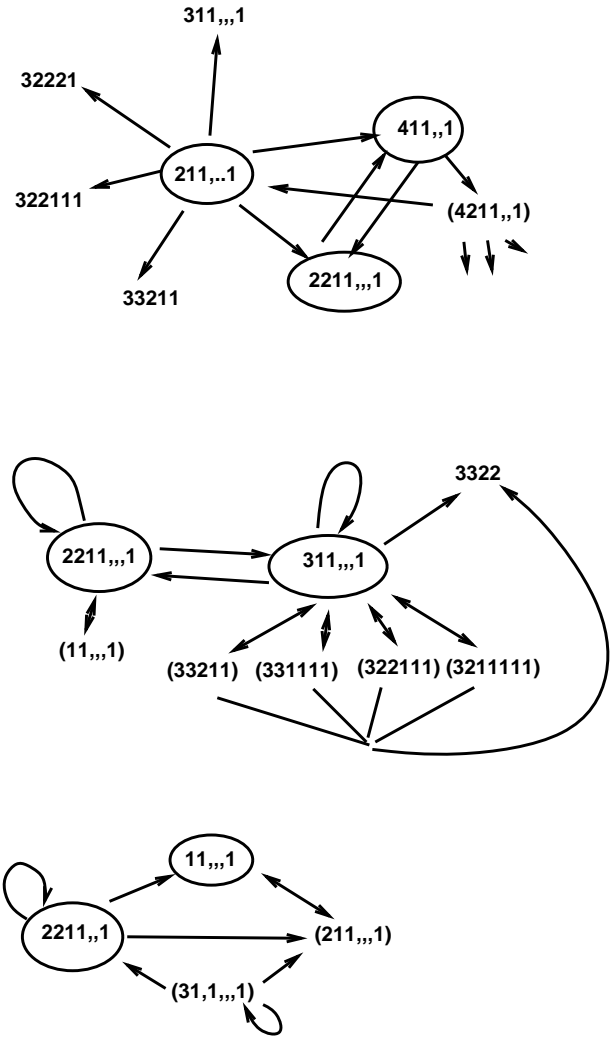,width=.6\textwidth}
\caption{
Examples of connection networks,
from $i$ to $j$, such that $T(i,j;+0) \neq 0$.
Connections that exists at $\sigma \rightarrow 0$ are plotted
for all fragile attractors, and most
psuedo attractors.  The arrow to itself is not the return to
itself (which always exists with some rate), but a switch to
a different attractor with the same clustering and different components.
The attractors enclosed in a circle are fragile attractors, and those
with the parenthesis are pseudo attractors, while those only with
the clustering are robust ones.
(a) $a=1.63$, (b)  $a=1.65$, and (c) $a=1.66$.}
\end{figure}
\vspace{.2in}


As $a$ is increased and the system enters the
PO phase, there appears connection network among fragile attractors,
as in Fig.19. Note that the arrow to itself
indicates not the return to the original ( since it occurs always with
some probability), but the transition from a different attractor
with the same clustering structure and with different 
components ( e.g., for the arrow to [2,1,1,...,1], the elements
forming a 2-element-cluster is different).

When $\sigma$ gets larger, there appears connection from some robust
attractors.  Note that the connection to fragile attractors 
is more frequent than the connection to robust attractors.  Hence,
in the presence of noise with larger $\sigma$, flow to fragile attractors 
may be larger than to robust attractors.
This will be important to noise-induced selection of Milnor attractors
in \S 9.

From several data in the connection matrix,
it may be possible to have the following picture on the phase space 
structure of our system:
In the ordered phase, several attractors exist far apart with each other.
The distance can be measured by the minimum perturbation to make the switch 
between the two.  At the complex ordered phase, 
several robust attractors still exist far apart, while
the fragile attractors exist in 
the intermediate region in the phase space, and
are connected to several robust attractors.
At the PO phase, basins of
Milnor attractors are often mutually connected.  Each Milnor 
attractor is connected with many other Milnor attractors, and
the connection is intermingled.
At the turbulent regime  the basin for a 
single attractor covers the whole phase space.


\section{Chaotic Itinerancy Revisited}

In high-dimensional dynamical systems, chaotic itinerancy among several 
ordered states is often observed\cite{KK-GCM,Tsuda,Ikeda}.
Orbits are globally attracted from a high-dimensional
chaotic state to these ordered states, where
they stay over long time steps, until
they exit from the state at a longer run.  These ordered states are
also called as attractor ruins, and are lower-dimensional objects in the phase
space.

One must note straightforwardly that the Milnor attractors satisfy 
the condition of the above ordered states constituting chaotic itinerancy.
When Milnor attractors that lose the stability ($P(0)<1$) keep attraction
for large $\sigma$, the total dynamics can be constructed as
the successive alternations to the attraction to them and
escapes from them.  Note that the Milnor attractor keeps
global attraction, which is consistent with the observation
that the attraction to ordered states in chaotic itinerancy 
occurs globally from a high-dimensional chaotic state.

The notion of chaotic itinerancy may be rather broad,
and some of CI may not
be explained by the Milnor attractor
network.  In particular, chaotic itinerancy in 
a Hamiltonian system\cite{Konishi,Shinjo}
may not fit directly with the present correspondence.  
Also, the `ordered states' in CI may not be close
enough to Milnor attractors.  Still,
the attribution of CI to Milnor attractor network dynamics is expected to
work as one ideal limit.



\section{Noise-induced Selection of Attractors}

Coexistence of attractors with different degrees of stability 
makes us expect the relevance of noise to the choice
of an attractor.  One might expect that the noise leads to the
choice of strong attractors.  To discuss this problem, we have 
simulated the model by applying a white noise with the amplitude $\delta$
(i.e., a random number homogeneously distributed over $[-\delta /2,\delta /2]$.
In Fig.20, we have plotted the temporal average of $\lambda_{spl}$ 
over all elements over 10000 steps.  Successive merging of
attractors is visible.  Here it should be noted that a robust attractor
is not necessarily selected.  In Fig.20a) in the order of
$\lambda_{spl}$ there are following attractors:
[2,1,..,1] (fragile; $\lambda_{spl}\approx .11$),
[3,1.,,.1](robust; $\lambda_{spl}\approx .024$),
[3,3,1,..,1](fragile $\lambda_{spl}\approx -0.09$), 
[4,4,1,1]( robust; $\lambda_{spl}\approx .-.17$), besides 
several robust attractors with $-.17<\lambda_{spl}<-.09$.
As seen in Fig. 20a), many orbits remain close to the
fragile attractors (around  $\lambda_{spl}\approx .11$ or $-0.09$),
even in the presence of noise.  On the contrary, robust attractors 
around  $-.17<\lambda_{spl}<-.09$
merge with a smaller strength of noise. 
In the example of Fig.20b), there are fragile
attractors 
[3,1,..,1](with the band (3:7); $\lambda_{spl}\approx .079$)
[3,1,..,1](with the band (4:6);  $\lambda_{spl}\approx .034$),
[2,2,1,..,1]{ $\lambda_{spl}\approx .02$), while the robust ones are
[2,2,2,2,1,1]($\lambda_{spl}\approx -.078$),
[3,3,2,2]( $\lambda_{spl}\approx -.095$), and
[3,2,2,1,1,1]($\lambda_{spl}\approx -.16$).  Again the fragile attractors
remain in the presence of noise.

\begin{figure}
\noindent
\epsfig{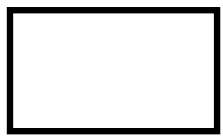}
\epsfig{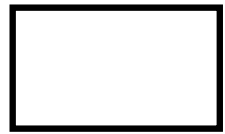}
\caption{
The splitting exponent $\lambda_{spl}$ averaged over 100000 time
steps, for a GCM with the noise term $(\delta /2)\times rnd_n(i)$
added throughout the temporal evolution.
Each dot represents
the value of $\lambda_{spl}$ from an initial condition, for
the corresponding noise amplitude value $\delta$ given by the horizontal axis,
while 1000 randomly chosen initial conditions are sampled for each value of
$\delta$.  (a) $a=1.62$.
(b) $a=1.64$.
}
\end{figure}
\vspace{.2in}


\begin{figure}
\noindent
\epsfig{file=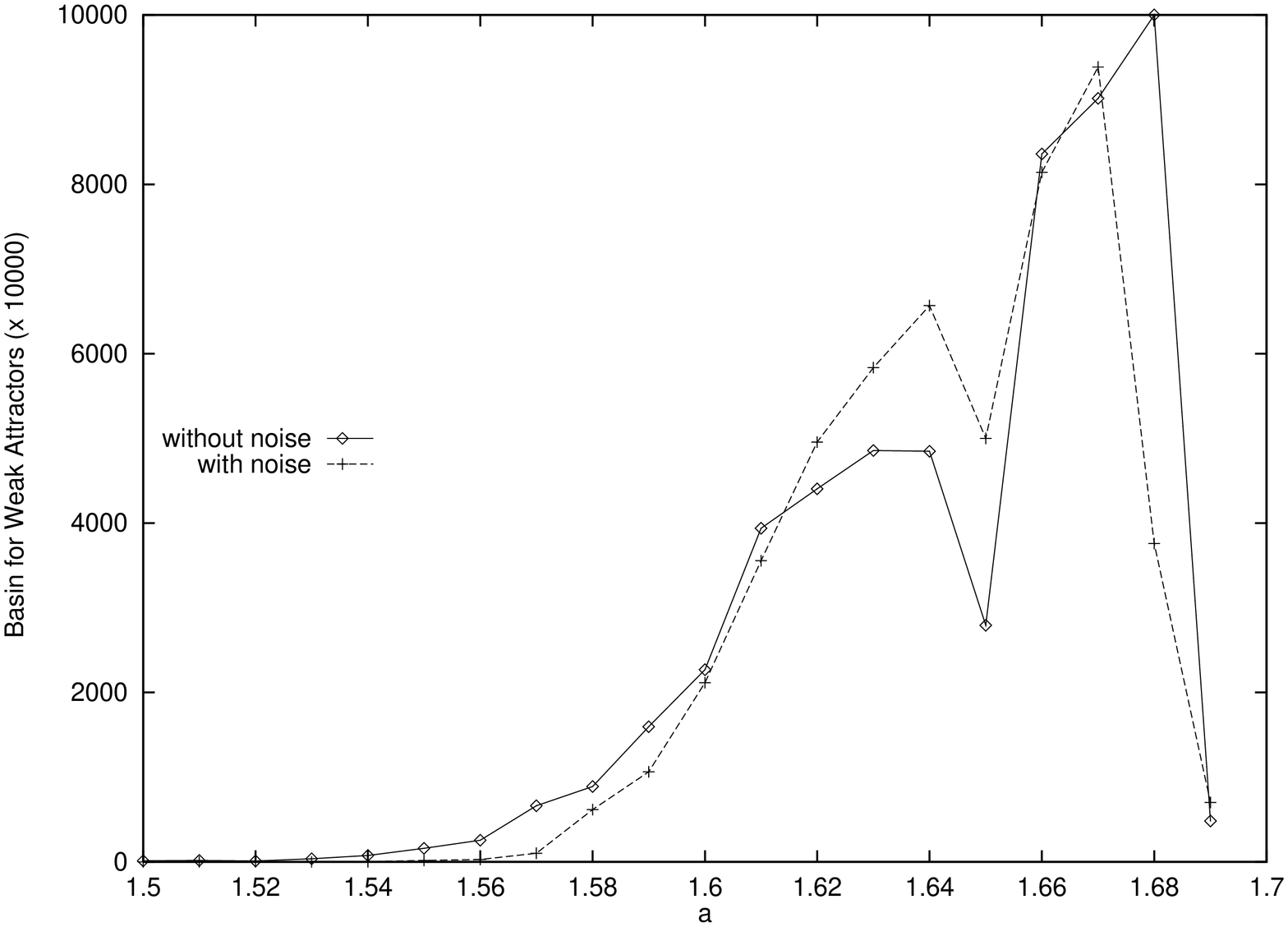,width=.6\textwidth,angle=-90}
\epsfig{file=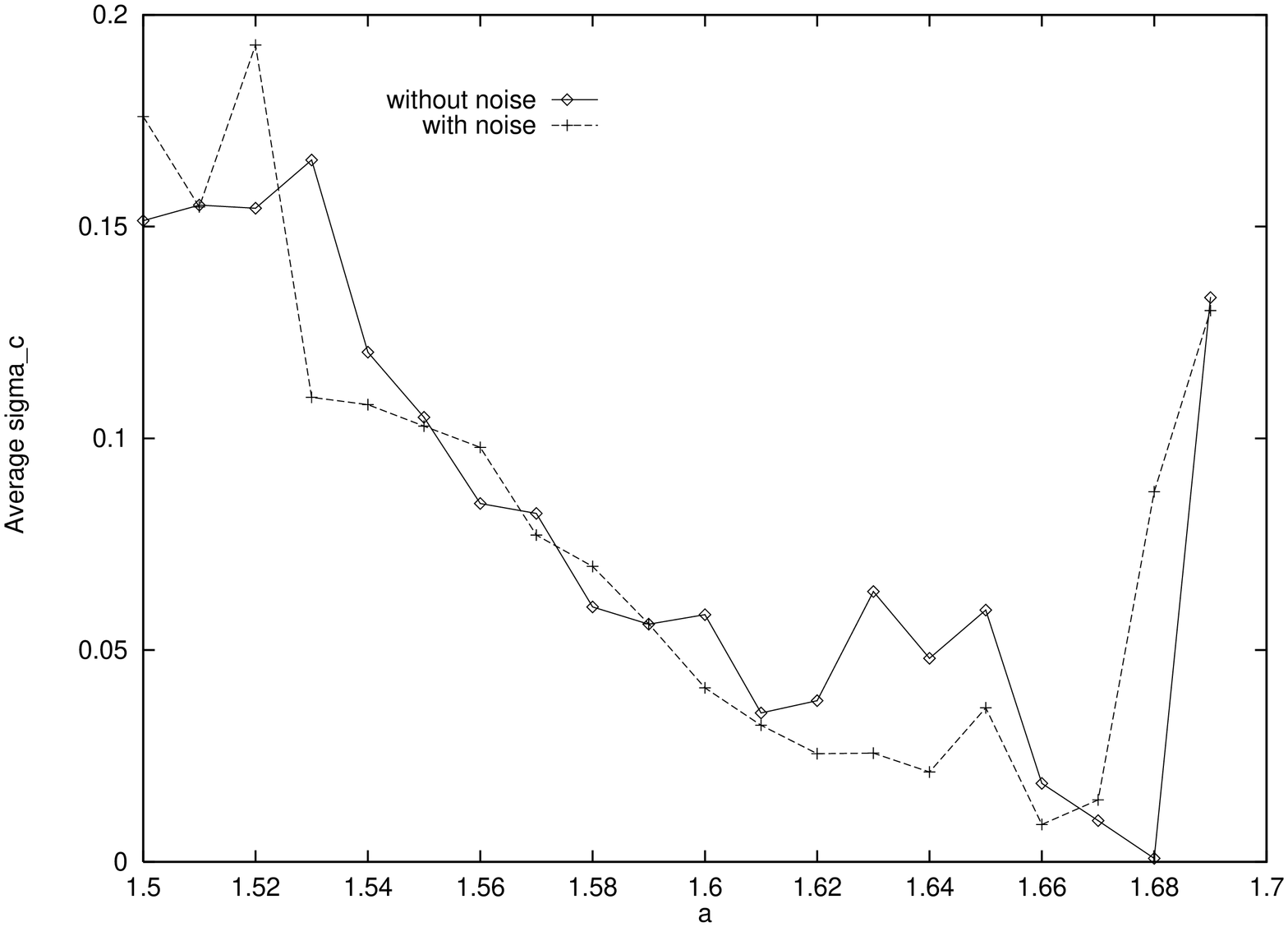,width=.6\textwidth,angle=-90}
\caption{
Change of the average strength $<\sigma_c>$ and
the rate of fragile attractors, versus the parameter value $a$.
Starting from random initial conditions, we have computed the GCM model
(1) with an additional noise term $(\delta/2) \times rnd_n(i)$ over $10^4$
steps with $\delta =0.01$ and checked which attractor is selected
after the noise is turned off.
$N$ is 25, although the same behaviors are seen for larger N.
(a)  the rate of fragile attractors with and without the noise term.
(b)  the average strength $<\sigma_c>$ ith and without the noise term.}
\end{figure}
\vspace{.2in}

When the noise is continuously added, however, the dynamics is represented by
successive switches over attractors.   Then, it is not easy to check 
the residence at each attractor in the presence of noise, since the criterion 
of  the neighborhood of each attractor is not clearly given.  Instead, we 
adopt a different method to check the noise effect, as
outlined in \S 2:  Iterate our dynamics in 
the presence of noise, over long enough time steps, and then turn off the 
noise, and check on which attractor the orbit falls.
With this process we measure the attraction ratio $V_i(\delta)$ for
a given attractor $i$.

Roughly speaking, our attraction rate gives an
estimate on  the residence time for the neighborhood of each attractor.
As it is very difficult to define the neighborhood of each attractor,
we use the present method as a numerically convenient tool.

Besides this convenience, this method is also relevant to consider the 
response of our system.
For example, in a neural system
some inputs are applied and then turned off.
Of course the input cannot be purely random, but consider
Freeman's experiment\cite{Freeman} for example, where
odor input is applied to a rabbit, and the neural activity in the olfactory
bulb is measured.  Long-term chaotic transient (of chaotic itinerancy
type) is observed for
unknown odor to the rabbit.
The external input of this odor,
may be regarded as almost random for a neural system of a rabbit.
In general the present method itself may be relevant to
the study of response of a nonlinear system against inputs.

Before discussing $V_i(\delta)$ for each attractor, we survey some
quantities over all attractors, averaged by the probability  $V_i(\delta)$
(i.e., average over initial conditions).
First our  numerical data show
that the average cluster number is decreased for the
ordered and partially ordered phases.  
In other words, there is a tendency that the synchronization among
elements is enhanced in the presence of noise.

This tendency leads to an opposite effect to the strength of attractors,
shown in Fig.21, where the
average of $\sigma_c$ over initial conditions and the
attraction rates to Milnor attractors are plotted, for $\delta=0.01$.
In the ordered (but not CO) and PO phases,
the attraction to robust attractors is slightly enhanced by the
noise, which leads to the increase of the average strength $<\sigma_c>$.
At the CO phase, however, the attraction rate to Milnor attractors
is increased by the noise, which leads to the
decrease of $<\sigma_c>$.  Dependence of the average
$<\sigma_c>$ on the noise amplitude $\delta$
is given in Fig.4 of ref.\cite{PRL}, where the decrease
is observed within some range of $\delta$
($0.04\stackrel{<}{\approx} \delta \stackrel{<}{\approx}0.2$), 
for the CO phase.  The mechanism of attraction 
to fragile attractors
is related with the robustness of global attraction, as will be discussed.

To see this mechanism in more detail, we have measured the dependence of
attraction rate to several attractors on the noise amplitude $\delta$.
A remarkable feature here
is its sensitivity in the choice of attractors on $\delta$.
At some noise strength,
attraction rate to some attractors is enhanced rather sharply.
After successive changes in the attraction rate,
it comes back to the level of noiseless case, for large $\delta$,
since, for large $\delta$, the memory of previous attractors is lost, which
essentially leads to random sampling of initial configurations. 

In Fig.22, we have plotted the number of initial points attracted to 
given attractors,
versus the noise strength applied during transient time steps.
For the ordered state ($a=1.5$), the noise has only a minor effect,
which just reduces the rate for an attractor with small $\sigma _c$.
On the other hand, in the CO phase,
there are successive enhancements of attraction rates to some attractors.
This stochastic amplification is a novel noise effect, which
reflects complex connection paths
among attractors.  The peak around $\delta \approx .04$ for
the attractor $[3,1,\cdots,1]$ and that around  $\delta \approx .05$
for the attractor $[4,1,\cdots,1]$ in Fig.22b), for example, are due to the
gap between the perturbation threshold allowing for the transitions 
from such attractors to others  and the reverse ones.
It should be noted that these attractors are fragile.

\begin{figure}
\noindent
\epsfig{file=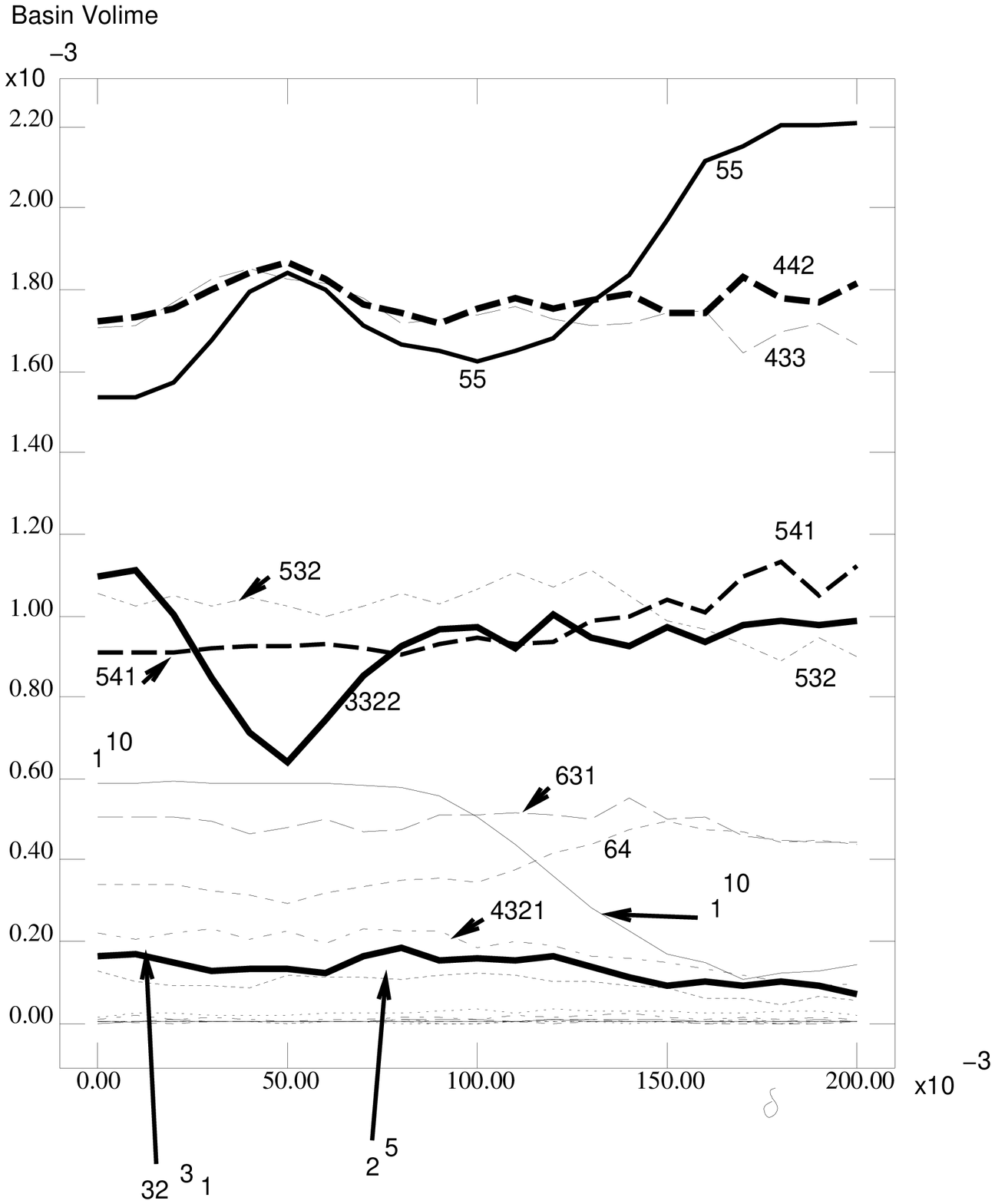,width=.6\textwidth}
\epsfig{file=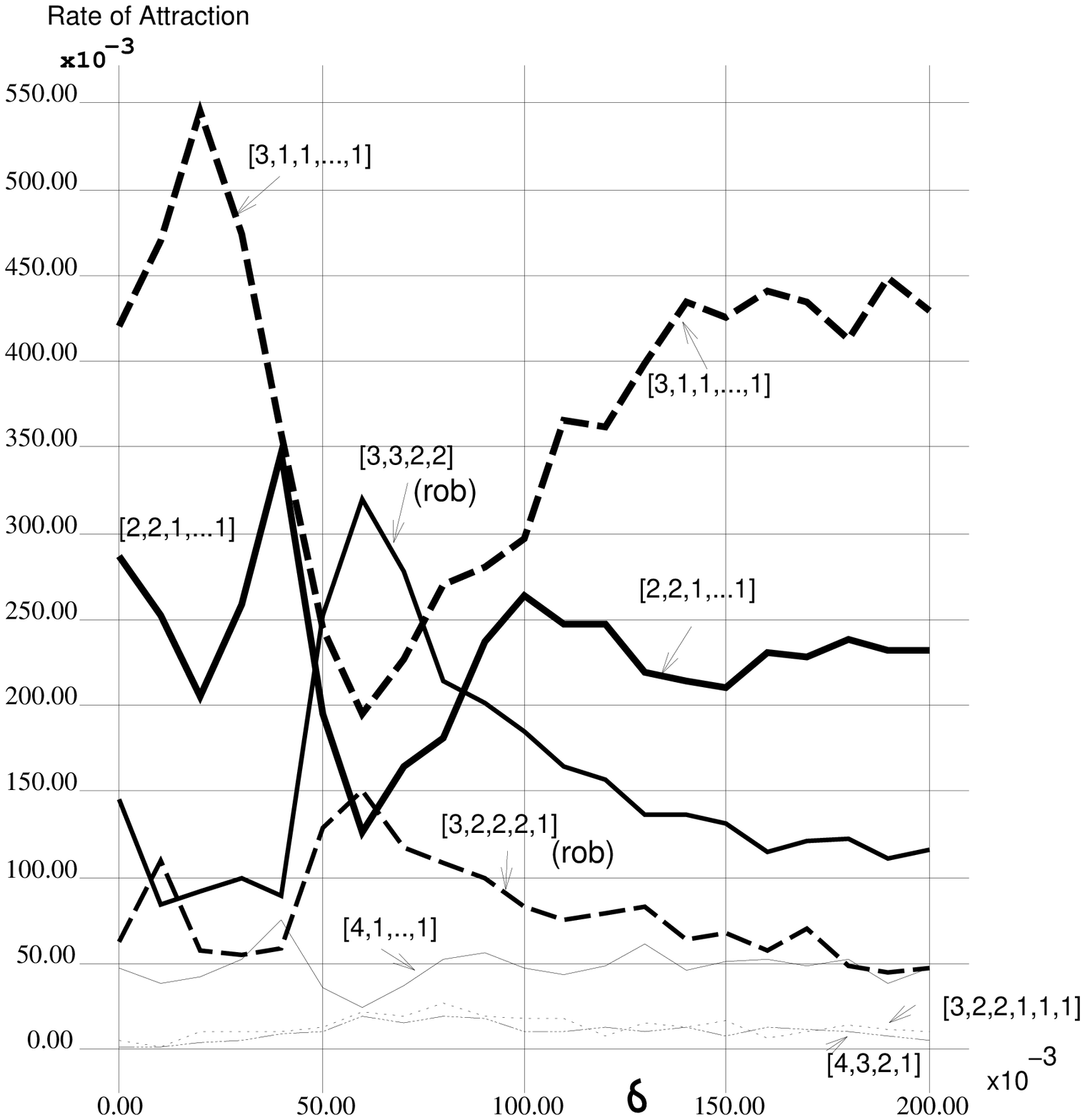,width=.6\textwidth}
\caption{
Rates of attraction to some attractors with the change of
transient noise amplitude $\delta$ for $N=10$.
Computations are carried out in the same manner as Fig.21.
(a) $a=1.5$ 
(b) $a=1.64$.
Only attractors with relatively large  attraction rates are plotted, where
those with ``rob" are robust attractors and others are fragile.}
\end{figure}
\vspace{.2in}

Why is the increase of the attraction to fragile attractors possible?
Although the detailed mechanism for it depends on the phase space structure,
it should be noted that there exists global attraction to fragile attractors,
as represented by $P(\sigma)$.  As is already  mentioned, 
fragile attractors often attract globally (for large
$\sigma$) more initial points than robust attractors.  Hence the orbits
kicked out of attractors may be attracted to fragile ( weak) attractors more.
When a large enough noise is added to kick
the orbit out of a robust attractor, the return rate to fragile attractors
can be larger than to robust ones.  Thus, when a noise amplitude exceeds
$\sigma_c$ of a robust attractor, the attraction rate to some fragile
attractors can increase.  Complicated structure in the attraction rate 
in Fig.22 reflects such
successive opening of the path from each robust attractor.


\section{Relevance to Biological Networks}

\subsection{Neural Dynamics: Dual Coding and Marginal Attractor}
 
It is interesting to note relevance of the present results to biological 
problems.  In neural network studies, dynamical systems with
global coupling is typically adopted,
although the coupling is not usually
identical.  Many features in GCM, however, are still valid even if the
coupling is not homogeneous.
Indeed, one can construct a chaotic neural network as
a globally coupled map with coded couplings, where the
partially ordered phase is relevant to the information 
processing \cite{Nozawa,NN}.

In Freeman's study mentioned earlier,
he has proposed that the chaotic dynamics
corresponds to a searching state for a variety of memories, represented by
attractors \cite{Freeman}.  Furthermore, Kay and Freeman have observed the
dynamics that can be regarded as chaotic itinerancy\cite{Kay}.

We note that the fragile attractors in the CO or PO phases provide a candidate 
for such a searching state, because of connection to a variety of
stronger attractors which possibly play the role of rigidly memorized states.
Selection of fragile attractors by some noisy inputs in \S 9 also
supports this correspondence.

It may be possible to introduce a degree of stability in memory,
corresponding to the degree of strength of attractors.
For a dynamical system
to work as a memory, some mechanism to write down and read it out is necessary.
If the memory is given in a robust attractor, its information processing
is not so easy, instead of its stability.  On the other hand,
Milnor attractors can support 
`dynamic memory'\cite{Tsuda,Tsuda?,Tsukada?,Edelman}.  
In a Milnor attractor,
some structure is preserved, while it
is dynamically connected with different attractors.
Also, it can be switched to different memory by any small inputs.
The connection to other attractors is neither one-to-one nor random.  It is
highly structurized with some constraints as discussed in \S 7,
while it keeps some variety.  
The switching process is expected to be hierarchically organized,
since the clustering in such attractors in the PO phase is hierarchical.
Hence the Milnor attractors
are good candidates for dynamic, hierarchical memory.
We propose that the Milnor attractor (network) is essential to the
interface process between inputs and robust memory, which is coded
by a robust attractor.

Another important feature in our system is dual coding. Note that
our attractor is coded by clustering condition.  Depending on 
possible combination of synchronization between elements, there
are a variety of attractors.  This coding by synchronization may
remind us of recently popular hypothesis on the temporal coding
\cite{Malsburg,Singer}
or dynamical cell assembly hypothesis\cite{Fujii}.  Against this type of
hypothesis, there are some criticisms pointing out
that, (i) a large number of
connections among units may be required, 
(ii) synchronization or de-synchronization may 
require long time, and that  (iii) another unit ( neuron) may be
necessary to detect synchronization \cite{Kawato}. It is interesting to 
mention that our system is free from all these criticisms.
First, only the connection to a single mean field is necessary in our system,
and we do not need $N\times N$ connections.  Second, the synchronization
and de-synchronization occur within a few time steps when an input
is applied to change the orbit.  Last, and most importantly, our system
has dual coding to overcome the third criticism.  Depending on
the way of synchronization, the dynamics of the mean field $h_n$ varies
( recall Fig.2 and Fig.3).  Instead of the condition for
synchronization (clustering), each attractor can be characterized by
a different type of mean-field dynamics (e.g., periodic or chaotic, 
the period of the cycle etc.).  It is important to note
that the check of synchronization requires comparison between
$N\times (N-1)/2$ pairs, while
the mean field has just a single variable.  Since this mean field is
applied to all elements, all elements ``know" to which type of attractor
they belong.  Thus the information on synchronization is also stored 
on each element.




\subsection{Relevance to Cell Biology}

Another possible application of our results is to cell biology.
In the context of dynamical systems it is sometimes assumed that each
cell type corresponds to an attractor of
some  internal cellular dynamics or genetic networks\cite{Kauffman}, while
the differentiation is
related with the selection of attractors. 

On the other hand, interference between internal cellular dynamics
and cell-to-cell interaction has explicitly been taken into account
in recent studies\cite{Cell,Cell2}. We have studied
a class of models with nonlinear intra-cellular dynamics, 
cell-to-cell interaction, and cell division to
increase the number.  It is found that  cells at an earlier
stage change their character by generation,
while the same character is preserved to offspring cells
at later generations.  As to the internal cellular dynamics,
this process is understood as a switch from a
weak attractor\footnotemark at the initial stage to a strong attractor later,
due to the cell-to-cell interaction.
When chaotic dynamics is allowed for internal dynamics, 
another type of cells is found that replicates
or switches to different types probabilistically \cite{CFKK}.  
This type corresponds to
stem cells.  Here the switched states keep their type by division, and are
regarded as determined cell types.
In the phase space, the intra-cellular dynamics of the stem-like cell has 
a wandering orbit visiting the neighborhoods of a few states that
correspond to determined cells.

\footnotetext{To be precise the corresponding dynamical state
is not necessarily an attractor, but often is a state
stabilized by the cell-to-cell interaction.}

Switch among attractors by a noise in \S 9 is important in this respect.
A Milnor attractor can switch to several robust attractors there.
A cell state, if represented by a Milnor attractor,
can switch to several different states depending on the interaction,
instead of the noise.  Noise-induced attraction to Milnor attractors
in \S 9 will be relevant to the appearance of stem cells.
It is also interesting to note that due to the chaotic dynamics therein,
the switch looks probabilistic as is often assumed in the
differentiation from a stem cell \cite{Ogawa}.
                      

\section{Summary and Discussion}

In the present paper, we have studied several aspects on the
strength of attractors in a high-dimensional dynamical system.

We have introduced the return probability of orbits
to an attractor, as a function of perturbation strength $\sigma$.
This function $P(\sigma)$ characterizes geometry of attraction 
to an attractor.  By introducing several quantifiers on the stability
of an attractor, it is found that the fragile (Milnor) attractors dominate
the basin volume in the partially ordered phase.

This dominance originates in the discrepancy between local stability
 and global attraction.  Milnor attractors which lose the local stability 
often keep global attraction.  Indeed, it is found that the global attraction is
rather robust against the change of bifurcation parameter, in contrast
with the local stability.  By the global attraction,
fragile attractors often have
relatively large basin volumes.

At some parameter regime in the PO phase, only Milnor attractors are observed,
where the dynamics is represented by switching process over  
Milnor attractor network.  Chaotic itinerancy, universally observed
as a high-dimensional, higher-level dynamics over 
low dimensional ordered states,
is re-interpreted as such Milnor attractor network dynamics.

Gap between local stability and global attraction also leads to a
rather strange noise effect.  Attraction rate to each attractor 
depends strongly on the noise amplitude.  Some attractors are
selectively attracted only with some range of noise amplitudes.
Furthermore,
Milnor attractors are selected for some range,
due to their global attraction.

Such selective attraction will be relevant to function in a biological
system.  Enzymatic activity of a biopolymer, for example, 
has a sharp dependence on  
temperature.  When the polymer dynamics is represented by
a high-dimensional dynamical system,
there should be a mechanism so that it responds selectively
to the amplitude of external noise.  Our model can provide an example
of such noise-selectivity,  where by switching the noise on and off,
it is possible to make a cyclic process between Milnor attractors and
robust attractors.\footnotemark

\footnotetext{Indeed, when a  coupled pendulum with many degrees of
freedom is under 
a heat bath and corresponding damping term, chaotic itinerancy is
observed, which allows for continuous energy absorption and storage\cite{KKNN}.}

Although our results are based on the GCM (1), it is expected
that the same qualitative behavior is observed 
in high-dimensional dynamical systems\cite{Hata}, 
since the previous findings in GCM\cite{KK-GCM,Wiesen} have been confirmed
in a coupled differential equations also\cite{JJ,Nakagawa}. 

One remaining question is the relevance of our results to a heterogeneous 
system.  We have adopted a GCM with identical elements, which makes us
easy to code an attractor only by clusterings.  If the elements are not
identical, complete synchronization between two elements is not possible.
Hence we have to check an attractor not by the condition $x(i)=x(j)$ but
by introducing the average ``distance" between $x_n(i)$ and $x_n(j)$.
Since the classification by the distance is not
automatic, we have to judge it case by case.  This is the reason
why we have treated only the homogeneous case.
Still, it is already verified that many attractors coexist in the
heterogeneous case\cite{Rel}.  Chaotic itinerancy dynamics is also seen in some
parameter region corresponding to the partially ordered phase.
We expect that our observation on the dominance and global attraction
of Milnor attractors, and the existence of
a Milnor attractor network are
valid in the heterogeneous case also.

Dominance of Milnor attractors gives us a suspect on the computability of
our system.  As long as digital computation is adopted,
it is always possible that an orbit is trapped to a state from which
it should depart by computation with a higher precision.
In this sense a serious problem is cast in numerical computation
of GCM\footnotemark.   

\footnotetext{Indeed, in our simulations 
we have often added a random floating at the
smallest bit of $x(i)$ in the the computer, to 
partially avoid such computational problem.}

This computation problem  also exists in the switching over Milnor attractor 
networks. In each event of switching, which Milnor attractor is visited next
after the departure from a Milnor attractor may 
depend on the precision.
In this sense the order of visits to Milnor attractors in chaotic itinerancy
may not be undecidable in a digital computation.
In other words, analog computation with GCM may decide what a digital
machine cannot do.  With this respect, it may be interesting to note
that there is a common statistical feature between (Milnor attractor) 
dynamics with a riddled basin 
and a Turing-machine dynamics having undecidability \cite{Saito-KK}. 

Existence of Milnor attractors may lead us to suspect the
correspondence between a (robust) attractor and memory,
often adopted in neuroscience (and theoretical cell biology).
It should be mentioned that Milnor attractors can provide
dynamic memory \cite{Tsuda,Tsuda?,Tsukada?,Edelman} allowing for 
interface between outside and inside, external inputs and internal
representation.

\vspace*{.2in}

I am grateful to  Naoko Nakagawa, Tatsuo Yanagita, Takashi Ikegami, and
Ichiro Tsuda for useful discussions.
This work is partially supported by a Grant-in-Aid for Scientific
Research from the Ministry of Education, Science, and Culture
of Japan.

\pagebreak
 
{\em Figure Captions}

Fig.1  Schematic representation of strength of an attractor:
a) overdamped motion of $x$ in a potential $U(x)$
b) dynamics without a potential
\vspace{.2in}

\vspace{.2in}
Fig.2  Overlaid time series of $x_n(i)$ of the attractor with the
clustering [1....1], accompanied by the time series of the
mean field. $a=1.66,\epsilon=.1,$ and $ N=10$.  Two examples.
(a) the band splitting with 5:5  (b) band splitting with 7:3.

\vspace{.2in}
Fig.3  Overlaid time series of $x_n(i)$ of the attractor with the
clustering [3,1...,1], accompanied by the time series of the
mean field. $a=1.65,\epsilon=.1, N=10$.  Three examples.
(a) the band splitting with 4:6  (b) band splitting with 7:3.

\vspace{.2in}

Fig.4 Change of the partition code for existing attractors,
with the change of $a$.
The vertical axis gives the number $[N_1N_2\cdots N_k]$.  For example 
the largest partition code corresponds to 1111111111, and the 
smallest one is 55.  By taking 
10000 initial conditions, and iterating our dynamics over 100000 steps,
we have checked on which attractor the orbit falls.  Hereafter
the basin volume rate is computed with this procedure, and
is measured as the sum of all rates over the
attractors with the same partition $[N_1,..N_k]$, unless otherwise mentioned.
The rate of initial conditions leading to such partition
is plotted as different marks.
$\triangle$( $>50$\%),$\times$ ($>10$\%),
$\Box$($> 5$\%), $+$($>1$\%), and $\Diamond$($>.1$\%).

\vspace{.2in}
Fig.5 The basin for each $Y$ value, with the change of $a$.
The rate of initial conditions leading to such value of $Y$
is plotted as different marks.
$\triangle$( $>50$\%),$\times$ ($>10$\%),
$\Box$($> 5$\%), $+$($>1$\%), and $\Diamond$($>.1$\%).
See text for the definition of $Y$.

\vspace{.2in}

Fig.6 Basin volume for attractors with $\lambda_{spl}$, as $a$ is changed.
The rate of initial conditions leading to such value of $\lambda_{spl}$
with the bin size 0.01 is plotted as different marks.
$\triangle$( $>50$\%),$\times$ ($>10$\%),
$\Box$($> 5$\%), $+$($>1$\%), and $\Diamond$($>.1$\%).
The average over 10000 initial conditions is also plotted as a line.
\vspace{.2in}

Fig.7 Change of the basin volume rates with the parameter $a$. $N=10$.
(a) the basin volume rates for two-cluster attractors.
(b) the basin volume rates for three-cluster attractors.
(c) the basin volume rates for attractors with the partition
$[\ell,1,\cdots,1]$.

\vspace{.2in}

Fig.8:  $P(\sigma)$ for several attractors
for $a=1.64$, and $N=10$.  For all the
figures,
we have 10000 initial conditions randomly chosen over $[-1,1]$ 
for each parameter, to make samplings. 
$P(\sigma$) is estimated by sampling over 1000 possible perturbations
for each $\sigma$. 
We often use the abbreviated
notation like $31^7$ for $[3,1,1,1,1,1,1,1]$.  Plotted are
robust attractors [32221] (with the basin volume rate $V \approx 6.3\%$ and  
$\sigma_c \approx .01$) and
[3322] (with $V \approx 15 $\% and  $\sigma_c \approx $.0012),
fragile attractors [31111111] (with $V \approx 42 $\%), [22111111]
(with $V \approx 29$ \%), and pseudo attractors [4321] (with $V \approx.2 $\%)
and [4111111] (with $V \approx 4.8 \%$).

\vspace{.2in}

Fig. 9:  $P(\sigma)$ for two-cluster attractors, with the clustering
[5,5], [6,4] [7,3], and [8,2].
\vspace{.2in}

Fig. 10:
 a)   $P(\sigma)$ for [2,1,1,1,...1] with the change of $a$
for $a=1,6,1.62,\cdots 1.63$.  The attractor is robust for $a=1.6$ and 1.61,
fragile for 1.62, and pseudo for 1.63.  The inlet is the expansion 
near $P(\sigma)=1$.

b) $P(\sigma)$ for [5,3,2] with the change of $a$.  The attractor is
fragile ($a=1.6$), pseudo (1.61) and robust (1.62).


\vspace{.2in}

Fig. 11:  
Strength $\sigma_c$ versus basin volume.  $\sigma_c$ is estimated
from $P(\sigma)$ measured by changing $\sigma$ 20\% successively 
from $10^{-5}$.  The points at $\sigma_c =10^{-5}$ just represent that
$\sigma_c <10^{-5}$. 
For Fig.11-13, we have estimated $\sigma_c$ from 100
possible perturbations: $\sigma_c$ is regarded to be less than the value
of $\sigma$ adopted in the run,
as long as all of 100 trials  result in the return to the
original attractor.

\vspace{.2in}

Fig. 12  
Dependence of $\sigma_c$ on the parameter $a$, for $N=100$.
By measuring $\sigma _c$ for attractors fallen from $10^4$ random initial
conditions, a histogram of
$log_{10}\sigma_c$ is constructed with a bin size 0.1.
The number of initial conditions leading to $log_{10}\sigma_c$ within the bin
is plotted as different marks.
$\triangle$( $>50$\%),$\times$ ( $>10$\%),
$\Box$($> 5$\%), $+$($>1$\%), and $\Diamond$($>.1$\%).
For all figures we have estimated $\sigma_c$ 
following the procedure given in the caption of Fig.11.
The points at $\sigma_c <10^{-4}$ just represent that
$\sigma_c <10^{-4}$.

\vspace{.2in}

Fig. 13:  Average strength $\sigma_c$ 
plotted as a function of $a$.  The basin volume of each
attractor is estimated as the rate of initial points leading to
the attractor, divided by the degeneracy.
The average is taken over $10^4$ random initial conditions.
For $N=10$, the strength is measured by increasing $a$ by 0.001,
while for $N=50$ and $N=100$ it is measured by the increment 0.01.

\vspace{.2in}
Fig.14:
The basin volume ratio of Milnor attractors with the change of $a$.
For each $a$, we take 1000 initial conditions, and iterate the
dynamics over 100000 steps to get an attractor.  We check if the
orbit returns to the original attractor,
by perturbing each attractor by $\sigma=10^{-7}$ over 100 trails. If the orbit
does not return at least for one of the trails, the
attractor is counted as a Milnor one.  For $N=10$, the ratio is
measured for $1.5<a<1.7$ with the increment 0.001, while
for larger sizes it is measured only for $1.62<a<1.7$ with the
increment 0.01.

\vspace{.2in}

Fig.15:  Change of the strength and basin volume rate with
the increase of $a$ by 0.001.  For each attractor,
the orbit is perturbed by $\sigma=10^{-7}$ over 1000 trails, to get
the return rate $P(+0)$, while the basin volume is measured
from $10^4$ initial conditions.  If none of the initial conditions
leads to the attractor, the strength is not plotted
(while the basin volume is plotted as 0).
(a) the attractor [3,1,..,1] with the band splitting (3:7)
(b) the attractor [2,2,1,..,1]
(c) the attractor [2,1,...,1]
(d)  the attractors [1,1,..,1] with the band splitting (10:0) 
(written as $1^{10}-0$), (7:3) ( $1^{10}-3$),
(6:4) ( $1^{10}-4$), and (5:5) ( $1^{10}-5$).
"S" shows the strength $P(0)$, "B" the basin volume ratio.

\vspace{.2in}

Fig.16
Size dependence of $<\sigma _m>$. The value
$\sigma_m$ is estimated from
$P(\sigma)$, measured by changing
$\sigma$ as $10^{-4+j/4}$ for ($j=0,1,\cdots,16$), by taking
1000 possible perturbations, while,
for $N=50$ and $N=100$, we have made only 100 possible perturbations.

\vspace{.2in}

Fig. 17:  $P(\sigma)$ for two-cluster attractors with the equal
partition.  $N=10$ ([5,5]), $N=20$,[10,10], $N=40$, $N=70$, and  $N=100$
([50,50]). $a=1.5$.

\vspace{.2in}

Fig. 18:
$P(\sigma)$ for several attractors with $a=1.65$.
(a) $N=50$ and  (b) $N=100$.

\vspace{.2in}

Fig.19: Examples of connection networks,  
from $i$ to $j$, such that $T(i,j;+0) \neq 0$.
Connections that exists at $\sigma \rightarrow 0$ are plotted
for all fragile attractors, and most
psuedo attractors.  The arrow to itself is not the return to
itself (which always exists with some rate), but a switch to
a different attractor with the same clustering and different components.
The attractors enclosed in a circle are fragile attractors, and those
with the parenthesis are pseudo attractors, while those only with
the clustering are robust ones.

(a) $a=1.63$, (b)  $a=1.65$, and (c) $a=1.66$,
\vspace{.2in}

Fig.20: The splitting exponent $\lambda_{spl}$ averaged over 100000 time 
steps, for a GCM with the noise term $(\delta /2)\times rnd_n(i)$ 
added throughout the temporal evolution.
Each dot represents 
the value of $\lambda_{spl}$ from an initial condition, for
the corresponding noise amplitude value $\delta$ given by the horizontal axis,
while 1000 randomly chosen initial conditions are sampled for each value of
$\delta$.  (a) $a=1.62$.  
(b) $a=1.64$. 
\vspace{.2in}

\vspace{.2in}
Fig.21: Change of the average strength $<\sigma_c>$ and
the rate of fragile attractors, versus the parameter value $a$.
Starting from random initial conditions, we have computed the GCM model
(1) with an additional noise term $(\delta/2) \times rnd_n(i)$ over $10^4$
steps with $\delta =0.01$ and checked which attractor is selected 
after the noise is turned off.
$N$ is 25, although the same behaviors are seen for larger N.
(a)  the rate of fragile attractors with and without the noise term.
(b)  the average strength $<\sigma_c>$ ith and without the noise term.
\vspace{.2in}

Fig.22: 
Rates of attraction to some attractors with the change of
transient noise amplitude $\delta$ for $N=10$.
Computations are carried out in the same manner as Fig.21.
(a) $a=1.5$ 
(b) $a=1.64$.
Only attractors with relatively large  attraction rates are plotted, where
those with ``rob" are robust attractors and others are fragile.


\begin{thebibliography}{999}
\bibitem{boundary}
The boundary of these phases change with $N$ when $N$ is not large enough 
( e.g.  $\stackrel{<}{\sim}$ 50).  For example, the region of PO phase
shifts upwards with the increase of $N$.  For $N =50$ and $100$,
the PO phase exists around 
$1.665\stackrel{<}{\approx}a\stackrel{<}{\approx}1.695$.
\end{thebibliography}
\end{document}